\documentclass{pnastwo}

\usepackage[dvips]{graphicx}
\usepackage{relsize}
\usepackage[tight,footnotesize]{subfigure}
\usepackage{mathrsfs}
\usepackage{longtable}
\usepackage{url}

\usepackage{verbatim}
\usepackage{xspace}
\usepackage{float}
\usepackage{perpage}
\usepackage{morefloats}

\usepackage[english]{babel} 
\usepackage{multirow}
\usepackage{dcolumn}
\usepackage{bm}
\usepackage{listings}
\usepackage{lineno}
\usepackage{color}

\usepackage{lmodern}

\graphicspath{{figures/}}

\usepackage{amssymb,amsfonts,amsmath}

\contributor{~~}
\copyrightyear{~~}
\issuedate{~~}
\volume{~~}
\issuenumber{~~}

\begin{document}


\title{Multi-particle field operators in quantum field theory}





\author{
Yu.V.~Volkotrub\affil{1}{Department of Theoretical and Experimental Nuclear Physics, Odessa National Polytechnic University, Shevchenko av. 1, Odessa, 65044, Ukraine},
M.A.~Deliyergiyev\affil{2}{Department of High Energy Nuclear Physics, Institute of Modern Physics,Nanchang Road 509, 730000 Lanzhou, China}
K.K.~Merkotan\affil{1}{Department of Theoretical and Experimental Nuclear Physics, Odessa National Polytechnic University, Shevchenko av. 1, Odessa, 65044, Ukraine},
N.A.~Chudak\affil{1}{Department of Theoretical and Experimental Nuclear Physics, Odessa National Polytechnic University, Shevchenko av. 1, Odessa, 65044, Ukraine},
O.S.~Potiyenko\affil{1}{Department of Theoretical and Experimental Nuclear Physics, Odessa National Polytechnic University, Shevchenko av. 1, Odessa, 65044, Ukraine},
D.A.~Ptashynskyy\affil{1}{Department of Theoretical and Experimental Nuclear Physics, Odessa National Polytechnic University, Shevchenko av. 1, Odessa, 65044, Ukraine},
G.O.~Sokhrannyi\affil{3}{Department of Experimental Particle Physics, Jo\v{z}ef Stefan Institute, Jamova 39, SI-1000 Ljubljana, Slovenia},
\affil{1}{Odessa National Polytechnic University, Shevchenko av. 1, Odessa, 65044, Ukraine}
A.V.~Tykhonov\affil{4}{D\'{e}partement de physique nucl\'{e}aire et corpusculaire, Universit\'{e} de Gen\'{e}ve, CH-1211 Geneva 4, Switzerland},
Yu.V.~Shabatura\affil{1}{Department of Theoretical and Experimental Nuclear Physics, Odessa National Polytechnic University, Shevchenko av. 1, Odessa, 65044, Ukraine},
I.V.~Sharf\affil{1}{Department of Theoretical and Experimental Nuclear Physics, Odessa National Polytechnic University, Shevchenko av. 1, Odessa, 65044, Ukraine},
\and
V.D.~Rusov\affil{1}{Department of Theoretical and Experimental Nuclear Physics, Odessa National Polytechnic University, Shevchenko av. 1, Odessa, 65044, Ukraine},
}

\contributor{}

\maketitle

\begin{article}

\begin{abstract}
Using bi-spinor fields we write the pseudo-scalar and bi-spinor fields that are characterized by the field functions of coordinates of several particles, namely multi-particle fields. By applying the quantization procedure to these multi-particle fields, hadronic creation and annihilation operators have been obtained. 
Due to internal degrees of freedom of such hadron field, it can interact with gauge fields. This interaction is introduced by a standard derivative extension. The gauge field which was obtained in this way revealed to be a multi-particle field. We construct the dynamic equations for this multi-particle gauge field. It was shown that the solutions of these equations can describe the interaction of quarks inside hadrons and also the interaction of quarks in different hadrons via two-particle gluon field. Quanta of this two-gluon field can be considered as bound states of two gluons. These solutions describe confinement of quarks and gluons.
\end{abstract}






\section{Introduction}

As was shown in references \cite{SharfUJP:2011, cej, SharfP1_JModPhys:2011, SharfP2_JModPhys:2012, SharfP3_JModPhys:2012}, in the framework of perturbative $\phi^{3}$ theory, it was possible to reproduce the qualitative features of the hadron's total cross-section spectrum as a function of $\sqrt{s}$. A new mechanism has been proposed that may explain the origin of the asymptotic growth of the total cross-section of hadron-hadron scattering. In framework of those models also possible to recreate inclusive cross-section spectra within changes of energy, where spectrum has just one maximum point at low energies, which may be replaced by two symmetrical maximum with increase of energy \cite{Sharph:2015eka}. 
However, results of such models allow to describe experimental data only qualitatively.

To move to the quantitative level one should manage to do the same computations in QCD, which is much more realistic theory of processes that occur in hadron scattering with respect to above mentioned $\phi^{3}$ theory. On the one hand within QCD one may use the same computation methods and consider the same physical mechanisms as in \cite{Sharf_Conf:2012, Sharf_Conf:2013}. On the other hand, there is a significant problem, which makes impossible an automatic translation of the results, which were obtained using perturbation theory in $\phi^{3}$, to perturbative QCD theory. This is a well-known problem, the interactions occur between quarks with help of gluons, while initial and final states always show up as their bound states (colorless), namely hadrons. From a general point of view of perturbation theory this observation manifested in the fact that Feynman diagrams use only quark and gluon lines, but there are no hadron lines. If one will consider such diagrams for construction of the scattering amplitude, the energy-momentum conservation law will be imposed on four-momenta of quarks, not on four-momenta of hadrons, as in experiments. This is a consequence of the fact that the Hamiltonian of a system do not reach asymptotically the Hamiltonian of free particles, if one will consider quarks and gluons as constituent particles of such a system. Therefore, we are not able to ``turn on" and to ``turn off" the interactions as in the regular $\hat{S}$ $\textendash$ matrix approach, since this interaction is the very essence of the mechanism that ensures the existence of hadrons as bound states of constituents quarks.

As it seems to us this problem is not associated exclusively with the application of the perturbation theory, rather it is related to the fact that all existing field theories are formulated in terms of single-particle occupation numbers of preceding states \cite{Altland:2006}. While quark states inside hadrons are not single-particle and principally cannot be expressed through single-particle states, since, due to relativistic invariance each such single-particle state should be characterized by a certain value of the energy-momentum. In the end, we return to the above mentioned problem of application of the energy-momentum conservation law, regardless of the choice of method for describing the scattering process. 

Usually these difficulties are overcome by use of parton models \cite{Brodsky:1998, Strikman:2011, MarkusDiehl:2012}. However, these models are much more adapted to the calculation of the inclusive characteristics of scattering processes rather than to the complete description of those processes. In particular, the system of multi-parton distribution functions for hadrons in the initial state, when there is no interaction between partons, is unknown. Determination of these functions is the problem by itself which yet further complicates the description of the scattering process.

However, if we believe that a hadron in the initial and final state of the scattering process, i.e. before or after interaction with other particles, is composed of a certain number of specific constituent quarks \cite{GellMann1964214, Zweig:570209, PhysRevLett.13.598, PhysRev.139.B1006}, then one may try to describe the internal state of such hadron using a non-relativistic approximation. In other words, one may try to describe this internal state by the probability amplitude, which can be found as a solution of the Schr\"{o}dinger equation with respect to the selected potential. 
It can be shown that if considering the non-relativistic bound state, one turns to the Jacobi coordinates \cite{Merkurev:1993} and splits out the internal state from the system-center-of-mass with respect to the initial reference frame, then the transition to the new reference frame should not change the internal state, as was proved in \cite{SharfState:2014, SharfState:2013}.

In this paper we propose to consider multi-particle field operators which will modify the occupation numbers of the multi-particle hadron states. The interaction of the colorless hadrons with gauge fields in this case can be described through the quark's internal degrees of freedom.

The constituent quarks will be described by bi-spinor fields $\Psi_{s}(x)$, $\bar{\Psi}_{s}(x)$, $s=1,2,3,4,...$. The corresponding field functions of these fields are defined in Minkowski space, denoted by $M_{x}, x \in M_{x}$. The range of values of quark's field functions is a linear space where either the bi-spinor representation of the Lorentz group or its Dirac-conjugate representation is realized. These spaces are denoted by ${B}$ and $\bar{B}$ respectively. Here we introduce the following notation:
\begin{equation}
\begin{split}
\hat{\Psi }=\left( \begin{matrix}
{{\Psi }_{1}}  \\
{{\Psi }_{2}}  \\
{{\Psi }_{3}}  \\
{{\Psi }_{4}}  \\
\end{matrix} \right),\hat{\bar{\Psi }}={{\hat{\Psi }}^{\dagger }}{{\hat{\gamma }}^{0}},
\end{split}
\label{eq:poznachenna1}
\end{equation}
where $\hat{\gamma}^{0}$ is the Dirac matrix.

Let's consider the tensor product of the two $M \otimes M$, or three Minkowski spaces, $M \otimes M \otimes M$ (for meson and baryons respectively) and the tensor product $\bar{B} \otimes B$, or $B \otimes B \otimes B$. The representations of the Lorentz group are realized on the tensor products $\bar{B} \otimes B$ or $B \otimes B \otimes B$. By expanding these products into direct sums of the invariant subspaces with respect to these representations one may reduce these representations to a pseudo-scalar form in case of $\bar{B} \otimes B$,  or bi-spinor form in case of $\bar{B} \otimes B$, or $B \otimes B \otimes B$:
\begin{equation}
\begin{split}
& \bar{B}\otimes B=I\oplus \cdots ,B\otimes B\otimes B=B\oplus \cdots , \\ 
& \bar{B}\otimes \bar{B}\otimes \bar{B}=\bar{B}\oplus \cdots . \\ 
\end{split}
\label{eq:neprivodimije}
\end{equation}  
Here ${I}$ is the invariant subspace on which the pseudo-scalar representation may be realized; the ellipses denote the rest of invariant subspaces which are not sufficient in this case.

Further, we can consider mapping  $M \otimes M$ into ${I}$, which we denote by $\phi(x_{1}, x_{2})$ and $\phi^{\ast}(x_{1}, x_{2})$, where $x_{1}, x_{2} \in M$. The mapping of $M \otimes M \otimes M$ manifold into ${B}$ and $M \otimes M \otimes M$ into $\bar{B}$ is denoted through $\Psi_{s}(x_{1}, x_{2}, x_{3})$, and $\bar{\Psi}_{s}(x_{1}, x_{2}, x_{3})$ respectively, where
$s=1,2,3,4, \rm{and}~x_{1},x_{2},x_{3}\in M$. 

Afterwards, we want consider the non-relativistic approximation for the internal dynamics of the constituent quarks. In this approximation the field functions $\phi(x_{1}, x_{2})$, $\Psi_{s}(x_{1}, x_{2}, x_{3})$, $\bar{\Psi}_{s}(x_{1}, x_{2}, x_{3})$ have to become the probability amplitudes, which have to be computed for the same values of time coordinates for all particles with respect to the reference frame, where we do our measurements. These issues were discussed in detail in Ref.\cite{SharfState:2014, SharfState:2013}. Consequently, later in the paper we will examine fields only on a submanifold of the entire tensor product $M \otimes M$, or  $M \otimes M \otimes M$, which has the following boundaries $x_{1}^{0}=x_{2}^{0}$, or $x_{1}^{0}=x_{2}^{0}=x_{3}^{0}$. 
It is convenient to make a transition to this submanifold by introducing the 4-space Jacobi coordinates on these tensor products:
\begin{equation}
\begin{split}
&{{X}^{a}}=\frac{x_{1}^{a}+x_{2}^{a}}{2},~~~~~~~~~y_{1}^{a}=x_{2}^{a}-x_{1}^{a}, \\ 
&{{X}^{a}}=\frac{x_{1}^{a}+x_{2}^{a}+x_{3}^{a}}{3},~~y_{1}^{a}=x_{3}^{a}-\frac{x_{1}^{a}+x_{2}^{a}}{2},~~y_{2}^{a}=x_{2}^{a}-x_{1}^{a}, \\ 
\end{split}
\label{eq:Jacobi}
\end{equation} 
where $a=0,1,2,3$. Here the first line corresponds to two-quark systems, and the second line $\textendash$ to three-quark systems.

On the subset that we will consider in this paper within variables of Eq.(\ref{eq:Jacobi}) one may select the following condition:
\begin{equation}
\begin{split}
& y_{1}^{0}=0,\quad {\rm{or}} \quad {y}_{1}^{0}={y}_{2}^{0}=0.
\end{split}
\label{eq:Odnochasnist}	
\end{equation} 

The internal variables we will denote as
\begin{align}
\mathbf{y}_{i}=\left( y_{i}^{1},y_{i}^{2},y_{i}^{3} \right),~i=1,2,
\label{eq:Vnutrzmin}	
\end{align}    
the multi-particle fields we denote as $\phi(X, \mathbf{y}_{1})$, $\Psi_{s}(X, \mathbf{y}_{1}, \mathbf{y}_{2})$, $\bar{\Psi}_{s}(X, \mathbf{y}_{1}, \mathbf{y}_{2})$, where we use the notation
\begin{align}
X\equiv \left( \begin{matrix}
{{X}^{0}}  \\
{{X}^{1}}  \\
{{X}^{2}}  \\
{{X}^{3}}  \\
\end{matrix} \right).
\label{eq:X}	
\end{align}  

Of course, the conditions in Eq.(\ref{eq:Odnochasnist}) cannot be imposed in a Lorentz invariant way. That is, observers in different inertial systems, by imposing the conditions of Eq.(\ref{eq:Odnochasnist}), will mark out different submanifolds on the corresponding tensor products of the Minkowski spaces. However, as was discussed in detail in \cite{SharfState:2014, SharfState:2013} in case of a general Lorentz transformation (i.e. one which has boost and cannot be reduced to pure rotations) internal changes in the different inertial reference frames cannot be connected between each other neither via Lorentz transformations nor by using any other methods. In case of rotations each of them behaves like a regular three-dimensional vector. Speaking further about Lorentz transformations we will merely bear in mind a transformation that cannot be reduced to rotations. Instead, the expression in Eq.(\ref{eq:X}) is transformed as a contravariant four-vector under Lorentz transformations. Thus, as was shown in \cite{SharfState:2014, SharfState:2013}, the dependence of  multi-particle fields on internal variables is the same in different inertial reference frames. Therefore, considering, for instance, the quantity 
\begin{align}
A=\int{{{{\bar{\Psi }}}_{s}}}\left( X,\mathbf{y}_{1},\mathbf{y_{2}} \right){{\Psi }_{s}}\left( X,\mathbf{y}_{1},\mathbf{y_{2}} \right)d\mathbf{y}_{1}d\mathbf{y_{2}},
\label{eq:A}	
\end{align}
we get in another reference frame (the repetition of the index ${s}$ implies summation over this index):     
\begin{equation}
\begin{split}
A & =\int d \mathbf{y_{1}^{\prime}} d \mathbf{y_{2}^{\prime}}  {{{\bar{\Psi }}}_{{{s}_{1}}}}\left( X={{\Lambda }^{-1}} X^{\prime},\mathbf{y_{1}^{\prime}},\mathbf{y_{2}^{\prime}} \right)D_{{{s}_{1}}s}^{-1}\left( \Lambda  \right) \\ 
& \times {{D}_{s{{s}_{2}}}}\left( \Lambda  \right){{\Psi }_{{{s}_{2}}}}\left( X={{\Lambda }^{-1}}X^{\prime},\mathbf{y_{1}^{\prime}},\mathbf{y_{2}^{\prime}}\right). \\ 
\end{split}
\label{eq:A1}
\end{equation}
Here $\Lambda$ is a Lorentz transformation, $D_{ss_{2}}(\Lambda)$ and $D_{s_{1}s}^{-1}(\Lambda)$ are the corresponding elements of the $\Lambda$ matrix transformation in the bi-spinor representation of the Lorentz group and its inverse transformation, respectively. At the same time the $\mathbf{y_{1}^{\prime}}$, $\mathbf{y_{2}^{\prime}}$ do not couple to $\mathbf{y}_{1}$, $\mathbf{y_{2}}$.  Given that the integrands in Eq.(\ref{eq:A}) and Eq.(\ref{eq:A1}) are the same functions of the internal variables, namely the integrand in Eq.(\ref{eq:A}) is the same function of variables $\mathbf{y}_{1}$, $\mathbf{y_{2}}$  as is the integrand in Eq.\ref{eq:A1} with respect to the variables $\mathbf{y_{1}^{\prime}}$, $\mathbf{y_{2}^{\prime}}$. One may note that the integrals differ only by notations of the integration variables. Therefore, $A$ takes the same values in different inertial reference frames, similar to the $\bar{\Psi}_{s}(X){\Psi}_{s}(X)$ expression which is the Lorentz invariant in case of ordinary single-particle field.

Integration over $\mathbf{y}_{1}$, $\mathbf{y_{2}}$ becomes the analog of summation of color and flavor indices, which justifies the use of the term ``internal variables" for these variables, since they are not connected to Lorentz transformations.

The above reasoning leads to the conclusion that, even though one cannot decouple expression Eq.(\ref{eq:Odnochasnist}) in a Lorentz-invariant way, the end result can still be harmonized with the principle of relativity.

For imposition of the multi-particle fields using the described method further in this work, we propose the dynamical equations and construct the corresponding Lagrangians with constants of motions. Then, we will implement the quantization procedure. 
The transformation of the field operators with respect to spacetime shift, not the boost \cite{Bogolyubov:1980}, is played an important role in the interpretation of field operators as those that are modifying the occupation numbers of the single-particle states. We would like to emphasize this statement because due to this shift, see Eq.(\ref{eq:SpaceTimeShifts}), multi-particle fields are manifested. As seen from the definition of Eq.(\ref{eq:Jacobi}), the internal Jacobi coordinates stay unchanged, while the coordinates $X^{a}$ are modified similarly to the normal single-particle coordinates. As will be shown later, this leads to the fact that the energy-momentum of a system becomes the energy-momentum of two-particle or three-particle systems, via the multi-particle field operators, that appear as a result of quantization of the multi-particle fields. Thus, they can be interpreted as the hadron's creation and annihilation operators. Therefore, by considering a theory with such operators one obtains the conservation law especially for four-momenta of hadrons as it should be.

So far, for brevity we did not write out the internal indices corresponding to the color, ${c}$, and flavor, ${f}$, for single-particle quark fields. Henceforth, we will assume that the single-particle quark fields have appropriate degrees of freedom, and they transform with respect to the fundamental representation of the $SU_{c}(3)$ and $SU_{f}(3)$ groups respectively. As a result of the tensor multiplication of these fields, the obtained multi-particle fields will transform with respect to tensor products of this representation. These representations are implemented on a linear space of values of the tensor multi-particle field functions. Laying out this space into a direct sum of subspaces that are invariant with respect to all transformations of considered representation, we select the subspace on which a trivial representation of  $SU_{c}(3)$ group has been implemented. Then we will consider a multi-particle field as an element of this subspace. In this way we take care of the colorlessness of hadrons. 
However, the presence of internal indices of the field, which is a consequence of the existence of internal quark degrees of freedom, allows one to introduce the interaction with the gauge field through the conventional derivative extension procedure. The requirement of simultaneity, Eq.(\ref{eq:Odnochasnist}), leads to the fact that this field should also be considered as multi-particle field. The operators of this field, as will be shown further, correspond to the creation and annihilation operators of bound states of gluons, see Eq.(\ref{eq:commutatori}) and the comments below.

Using the multi-particle hadron fields that can interact by the multi-particle gauge fields, which in turn may generate secondary particles by interactions with the hadron fields, one may obtain a method to describe the processes of elastic and inelastic scattering of hadrons with the ``correct" law of energy-momentum conservation, which is imposed on the energy-momenta of hadrons rather than on constituent quarks.

\section{Two-particle scalar field}
\label{2particle_ScalarField}

As the simplest example model of multi-particle fields let us consider the two-particle scalar field.

Let us consider a system of two noninteracting scalar particles. Usually the state of such system is described by the Fock-space elements:
\begin{equation}
\begin{split}
\left| \Psi  \right\rangle =\left( \begin{matrix}
0  \\
0  \\
{{\Psi }_{2}}\left( t,\mathbf{r}_{1},\mathbf{r}_{2} \right)  \\
0  \\
\vdots   \\
\end{matrix} \right). 
\end{split}
\label{eq:Fok}
\end{equation} 
Dynamics of this state is described by the time evolution operator that is constructed on a basis of the single-particle Klein-Gordon-Fock equation \cite{Schrodinger:1926, OKlein:1926, Fock:1926, Gordon:1926}. However, due to the fact that we are considering the system of noninteracting particles, the two-particle state in process of dynamical evolution will remain a two-particle state. Moreover, we are going to include interactions between those particles to the system, even through the characteristic values of the energy of this interaction and the energy of the relative motion of these particles are negligible $\textendash$ i.e. not enough to produce another particles. The system with such interactions will remain a two-particle system even if one would like to consider it with respect to the reference frame where the entire system has relativistic energy. 
Thus, one can ask questions about the dynamical two-particle operator in this case.

On the other hand, the square modulus of $\Psi_{2}(t,\mathbf{r}_{1},\mathbf{r}_{2})$ is equal to the joined density probability for the coordinates of two particles, if measurements of the coordinates will be performed simultaneously with respect to the reference frame where the state Eq.(\ref{eq:Fok}) is considered. The dynamic operator, which is built by using the energy operator for the single-particle Klein-Gordon-Fock field ignores this simultaneity; therefore it also leads to the issue of the construction of the two-particle operator.

Given that we consider noninteracting particles, one may write the Klein-Gordon-Fock equation for each of them and examine the following system of equations:
\begin{equation}
\begin{cases} 
\scalebox{1.2}{$ -g^{ab}\frac{\partial^{2}\phi(x_{1}) }{\partial x_{1}^{a}\partial x_{1}^{b}} -m^{2}\phi(x_{1})=0,  $}\\ 
\scalebox{1.2}{$ -g^{ab}\frac{\partial^{2}\phi(x_{2}) }{\partial x_{2}^{a}\partial x_{2}^{b}} -m^{2}\phi(x_{2})=0. $}\\ 
\end{cases}
\label{eq:FockSystemEq}
\end{equation}
Note, that here and further the Klein-Gordon-Fock equation will be written using the Bogolyubov convention, where $g^{ab}$ is the Minkowski tensor. The value of functions $\phi(x_{1})$ and $\phi(x_{2})$ at the arbitrary points $x_{1}$ and $x_{2}$ are independent variables, so the system of equations, Eq.(\ref{eq:FockSystemEq}), is equivalent to the equation:
\begin{equation}
\begin{split}
& \left( -{{g}^{ab}}\frac{{{\partial }^{2}}\phi \left( {{x}_{1}} \right)}{\partial x_{1}^{a}\partial x_{1}^{b}}-{{m}^{2}}\phi \left( {{x}_{1}} \right) \right)\phi \left( {{x}_{2}} \right) \\ 
& +\phi \left( {{x}_{1}} \right)\left( -{{g}^{ab}}\frac{{{\partial }^{2}}\phi \left( {{x}_{2}} \right)}{\partial x_{2}^{a}\partial x_{2}^{b}}-{{m}^{2}}\phi \left( {{x}_{2}} \right) \right)=0. \\ 
\end{split}
\label{eq:2KGF}
\end{equation}
At this point we introduce the two-particle scalar field $\phi(x_{1}, x_{2})$:
\begin{equation}
\begin{split}
\phi \left( {{x}_{1}},{{x}_{2}} \right)=\phi \left( {{x}_{1}} \right)\phi \left( {{x}_{2}} \right).          
\end{split}
\label{eq:Dvo_chast}
\end{equation}
Then Eq.\ref{eq:2KGF} can be rewritten as:
\begin{equation}
\begin{split}
-{{g}^{ab}}\frac{{{\partial }^{2}}\phi \left( {{x}_{1}},{{x}_{2}} \right)}{\partial x_{1}^{a}\partial x_{1}^{b}} -{{g}^{ab}}\frac{{{\partial }^{2}}\phi \left( {{x}_{1}},{{x}_{2}} \right)}{\partial x_{2}^{a}\partial x_{2}^{b}} -2{{m}^{2}}\phi \left( {{x}_{1}},{{x}_{2}} \right)=0. \\   
\end{split}
\label{eq:2KGF1}
\end{equation}
Note, that writing the functions of $x_{1}$ under the sign of the derivative of $x_{2}$ and vise versa is possible everywhere in the domain where the function, Eq.(\ref{eq:Dvo_chast}), is defined; except for the subset where $x_{1}=x_{2}$. Along the surface $x_{1}=x_{2}$, this operation is forbidden. However, this subset has a zero measure in the whole tensor product of the two Minkowski spaces for two particles. Given that the equations which will be considered, do not lead to existence of certain features of the two-particle function, Eq.(\ref{eq:Dvo_chast}), on this subset, it should give us vanishing contribution to the integrals over the domain, where this two-particle function is defined. Thus, this subset can be ignored in the integrals that define the physical quantities. That is why, here and in the remainder of the paper, we are going to make analogical transformations.

Now we show, that if one will examine Eq.(\ref{eq:2KGF1}) as dynamical equations for two-particle scalar field $\phi(x_{1}, x_{2})$, then one will get a meaningful result from a physical point of view.

The Eq.(\ref{eq:2KGF1}) can be obtained as the Euler-Lagrange equation from Lagrangian:
\begin{equation}
\begin{split}
L\left( {{x}_{1}},{{x}_{2}} \right) & =\frac{1}{2}\left( {{g}^{ab}}\frac{\partial \phi \left( {{x}_{1}},{{x}_{2}} \right)}{\partial x_{1}^{a}}\frac{\partial \phi \left( {{x}_{1}},{{x}_{2}} \right)}{\partial x_{1}^{b}} \right. \\ 
& \left. +{{g}^{ab}}\frac{\partial \phi \left( {{x}_{1}},{{x}_{2}} \right)}{\partial x_{2}^{a}}\frac{\partial \phi \left( {{x}_{1}},{{x}_{2}} \right)}{\partial x_{2}^{b}} -2{{m}^{2}}{{\phi }^{2}}\left( {{x}_{1}},{{x}_{2}} \right) \right). \\   
\end{split}
\label{eq:Lagrangianfi2}
\end{equation}
Transforming this Lagrangian with the help of the two-particle four-space Jacobi coordinates \cite{SharfP1_JModPhys:2011} we get: 
\begin{equation}
\begin{split}
L\left( X,{{y}_{1}} \right) & =\frac{1}{2}{{g}^{ab}}\left( \frac{1}{2}\frac{\partial \phi \left( X,{{y}_{1}} \right)}{\partial {{X}^{a}}}\frac{\partial \phi \left( X,{{y}_{1}} \right)}{\partial {{X}^{b}}} \right. \\ 
& \left. +2\frac{\partial \phi \left( X,{{y}_{1}} \right)}{\partial y_{1}^{a}}\frac{\partial \phi \left( X,{{y}_{1}} \right)}{\partial y_{1}^{b}} -2{{m}^{2}}{{\phi }^{2}}\left( X,{{y}_{1}} \right) \right). \\   
\end{split}
\label{eq:Lagrangianfi2Jacobi}
\end{equation} 
Let consider the Lagrangian Eq.(\ref{eq:Lagrangianfi2Jacobi}) for the functions defined on the subset in Eq.(\ref{eq:Odnochasnist}). Due to the fact that the functions along this subset do not depend on $y_{1}^{0}$ the corresponding derivatives are equal to zero and the Lagrangian takes the form:
\begin{equation}
\begin{split}
L\left( X,\mathbf{y}_{1} \right)& =\left( \frac{1}{4}{{g}^{ab}}\frac{\partial \phi \left( X, \mathbf{y}_{1} \right)}{\partial {{X}^{a}}}\frac{\partial \phi \left( X, \mathbf{y}_{1} \right)}{\partial {{X}^{b}}} \right. \\ 
& -\sum\limits_{d=1}^{3}{\left( \frac{\partial \phi \left( X, \mathbf{y}_{1} \right)}{\partial y_{1}^{d}}\frac{\partial \phi \left( X,\mathbf{y}_{1} \right)}{\partial y_{1}^{d}} \right)}  \\ 
& \left. -2{{m}^{2}}{{\phi }^{2}}\left( X,\mathbf{y}_{1} \right) \right). \\    
\end{split}
\label{eq:Lfi2Odnochasnij}
\end{equation} 

Note, that from this moment on, the field $\phi(X, \mathbf{y})$, generally speaking, ceases to be a product of solutions of Eq.(\ref{eq:2KGF1}), because these equations, even if one will consider them at the same values of time $x_{1}^{0}=x_{2}^{0}$, contain derivatives $\partial/\partial x_{1}^{0}$ and $\partial/\partial x_{2}^{0}$ respectively (derivatives at ``their own" times). Whereas in consideration of the equations for the field $\phi(X, \mathbf{y})$ it is allowed just a derivative along the plane of the simultaneity defined as Eq.(\ref{eq:Odnochasnist}). Physically it means that we have a reduced symmetry with respect to the time shift. If, before the transition to the plane of the simultaneity, one may shift each of the variables $x_{1}^{0}$ and $x_{2}^{0}$ independently from each other, afterwards, once we move to this plane, one can shift only the common value of these variables. As a result, instead of the energies of the separate particles, we obtain the energy of the whole two-particle system. Formally, now we can say that instead of the two single-particle fields we obtain a new two-particle field, which is separate from them.

The action will be defined as an integral over the Lagrangian from Eq.(\ref{eq:Lfi2Odnochasnij}) on a selected submanifold:
\begin{equation}
\begin{split}
S=\int d^{4}X d\mathbf{y}_{1} L\left( X,\mathbf{y}_{1} \right).  
\end{split}
\label{eq:diaOdnochas}
\end{equation}

As was discussed in the previous section, if one will consider the components of $\mathbf{y}_{1}$ as internal variables, then this action has the same values in all inertial reference frames.

The Euler-Lagrange equation that is generated by the Lagrangian of  Eq.(\ref{eq:Lfi2Odnochasnij}) has the form:
\begin{equation}	\begin{split}
-{{g}^{cd}}\frac{{{\partial }^{2}}\phi \left( X, \mathbf{y}_{1} \right)}{\partial {{X}^{c}}\partial {{X}^{d}}}&-\left( {{\left( 2m \right)}^{2}}\phi \left( X, \mathbf{y}_{1} \right) \right. \\ 
& \left. -2\left( 2m \right)\left( \frac{1}{m}{{\Delta }_{\mathbf{y}_{1}}}\phi \left( X, \mathbf{y}_{1} \right) \right) \right)=0. \\ 
\end{split}
\label{eq:LagEiler}
\end{equation} 
Here by ${\Delta }_{\mathbf{y}_{1}}$ we denote the Laplace operator, which is defined with respect to the components of the vector $\mathbf{y}_{1}$.

If we consider the non-relativistic problem of two noninteracting particles then in terms of the Jacobi variables one will obtain the expression for the internal Hamiltonian of the system as:
\begin{equation}
\begin{split}
{{\hat{H}}^{\text{internal}}}=\left( 2m \right)\hat{E}-\frac{1}{m} \Delta_{\mathbf{y}_{1}},  
\end{split}
\label{eq:Hinternal}
\end{equation}
where $\hat{E}$ is a unity operator. In this non-relativistic approximation, the eigenvalues of the operator $\left( -\frac{1}{m} \Delta_{\mathbf{y}_{1}} \right)$ should be small with respect to $2m$. Therefore, with a precision up to order of $\left( -\frac{1}{m} \Delta_{\mathbf{y}_{1}} \right)^{2}$, instead of Eq.(\ref{eq:LagEiler}) we can write:
\begin{equation}
\begin{split}
& -{{g}^{cd}}\frac{{{\partial }^{2}}\phi \left( X, \mathbf{y}_{1} \right)}{\partial {{X}^{c}}\partial {{X}^{d}}}-{{\left( {{{\hat{H}}}^{\text{internal}}} \right)}^{2}}\phi \left( X,\mathbf{y}_{1} \right)=0. \\ 
\end{split}
\label{eq:LagEiler1}
\end{equation}

This equation is obtained for noninteracting fields. The internal Hamiltonian has the form of the kinetic energy operator defined with respect to the motion of free particles. In later sections it will be shown that if one uses the Lagrangian with extended derivatives instead of the free Lagrangian of the type in Eq.(\ref{eq:Lagrangianfi2}), then the internal Hamiltonian will include the corresponding operator of the potential energy, which will guarantee the existence of the bound state. For a stable bound particle, the dependence of the field $\phi(X, \mathbf{y}_{1})$, from internal variables can be determined by the requirement that this function has to be an eigenfunction of the internal Hamiltonian corresponding to the smallest eigenvalue. Then, instead of the mass-squared term in the ``original" Klein-Gordon-Fock equation, for the two-particle equation the square of this smallest eigenvalue will appear, as expected. 

In this section, we still continue the consideration of the two-particle scalar field, constructed from the two noninteracting fields. It is not difficult to implement the usual description program of quantum fields \cite{Bogolyubov:1980} for the two-particle field $\phi(X, \mathbf{y}_{1})$. In particular, the general solution of Eq.(\ref{eq:LagEiler}) can be written as a sum of the negative and positive frequency solutions $\phi(X, \mathbf{y}_{1})=\phi^{-}(X, \mathbf{y}_{1})+\phi^{+}(X, \mathbf{y}_{1})$, where:
\begin{equation}
\begin{split}
\phi^{-}(X, \mathbf{y}_{1}) & =\frac{1}{(2\pi)^{3/2}}\int{\frac{d\mathbf{p}d\mathbf{q}_{1}}{\sqrt{2{{p}_{0}}\left(\mathbf{p}, \mathbf{q}_{1} \right) }}} \\ 
& \times \phi^{-}\left(\mathbf{q}_{1}\cdot\mathbf{y}_{1} \right) \exp \left( -ipX-i\left( \mathbf{q}_{1}\cdot\mathbf{y}_{1} \right) \right), \\ 
\end{split}
\label{eq:Negativnochastotne}
\end{equation}
\begin{equation}
\begin{split}
\phi^{+}(X, \mathbf{y}_{1}) & =\frac{1}{(2\pi)^{3/2}}\int{\frac{d\mathbf{p}d\mathbf{q}_{1}}{\sqrt{2{{p}_{0}}\left(\mathbf{p}, \mathbf{q_{1}} \right) }}} \\ 
& \times \phi^{+}\left(\mathbf{q}_{1}\cdot\mathbf{y}_{1} \right) \exp \left( ipX+i\left( \mathbf{q}_{1}\cdot\mathbf{y}_{1} \right) \right), \\ 
\end{split}
\label{eq:Pozitivnochastotne}
\end{equation}
Here by ${p}$, we denote the four-vector:
\begin{equation}
\begin{split}
{{p}_{a}}=\left( \begin{matrix}
{{p}_{0}}\left( \mathbf{p}, \mathbf{q}_{1} \right)  \\
-{{p}_{x}}  \\
-{{p}_{y}}  \\
-{{p}_{z}}  \\
\end{matrix} \right).
\end{split}
\label{eq:scalar4imp}
\end{equation}
The zero component is defined from the mass shell requirement
\begin{equation}
\begin{split}
{{p}_{0}}\left( \mathbf{p}, \mathbf{q_{1}} \right)=\sqrt{{{\left(2m\right)}^{2}} + \mathbf{p}^{2} + \mathbf{q_{1}}^{2}}.  
\end{split}
\label{eq:UmovaMasPover}
\end{equation} 
 
For the multi-particle fields one may prove Noether's theorem and derive the constants of motion. Hereby, due to transformation of the spacetime shift that can be written as:
\begin{equation}
\begin{split}
{x^{\prime}}_{1}^{a}& = x_{1}^{a}+\Delta x^{a},\\
{x^{\prime}}_{2}^{a}& = x_{2}^{a}+\Delta x^{a},\\
\phi^{\prime}& = \phi
\end{split}
\label{eq:SpaceTimeShifts}
\end{equation}
as seen from Eqs.(\ref{eq:Jacobi}), nontrivial transformation has only $X^{a}$ while internal coordinates remain unchanged.

The energy-momentum four-vector of the multi-particle field with the help of Noether's theorem, can be written using the negative and positive frequency coefficients $\phi^{-}(\mathbf{p},\mathbf{q}_{1})$ and $\phi^{+}(\mathbf{p},\mathbf{q}_{1})$ that are included into the formulas Eq.(\ref{eq:Negativnochastotne})-(\ref{eq:Pozitivnochastotne}), in the usual for scalar field way, with the only exception that the internal integration over the internal momentum $\mathbf{q}_{1}$ is added:
\begin{equation}
\begin{split}
{{P}^{a}}=\int{d\mathbf{p}d\mathbf{q}_{1} {{p}^{a}}}{{\phi }^{+}}\left( \mathbf{p},\mathbf{q}_{1} \right){{\phi }^{-}}\left( \mathbf{p},\mathbf{q}_{1} \right).  
\end{split}
\label{eq:enimp}
\end{equation}
When the index ${a}$ is equal to 0, the $p^{0}$ in Eq.(\ref{eq:enimp}) is defined through formula (\ref{eq:UmovaMasPover}).

Due to the fact that during the transformation of the spacetime shift, only the components of the four-vector $X^{a}$ are changed, in the transition to the quantum field theory one may obtain 
in a conventional manner \cite{Bogolyubov:1980} that the field operators $\phi(X, \mathbf{y}_{1})$ fulfill the requirement:
\begin{equation}
\begin{split}
\frac{\partial \hat{\phi }\left( X,\mathbf{y}_{1} \right)}{\partial {{X}^{a}}}=i\left[ {{{\hat{P}}}_{a}},\hat{\phi }\left( X, \mathbf{y}_{1} \right) \right]  
\end{split}
\label{eq:Rivnanna}
\end{equation}    

As a consequence of this equation, taking into account Eq.(\ref{eq:Negativnochastotne}) and (\ref{eq:Pozitivnochastotne}), we got the following relations:
\begin{equation}
\begin{split}
\left[ {{{\hat{P}}}_{a}},{{{\hat{\phi }}}^{-}}\left( \mathbf{p},\mathbf{q}_{1} \right) \right]& =-{{p}_{a}}{{{\hat{\phi }}}^{-}}\left( \mathbf{p},\mathbf{q}_{1} \right), \\ 
\left[ {{{\hat{P}}}_{a}},{{{\hat{\phi }}}^{+}}\left( \mathbf{p},\mathbf{q}_{1} \right) \right]& ={{p}_{a}}{{{\hat{\phi }}}^{+}}\left( \mathbf{p},\mathbf{q}_{1} \right), \\ 
\end{split}
\label{eq:commutatori}
\end{equation}
which allow one to interpret $\hat{\phi}^{-}(\mathbf{p},\mathbf{q}_{1})$ and $\hat{\phi}^{+}(\mathbf{p},\mathbf{q}_{1})$ as creation and annihilation operators. Thus, based on Eq.(\ref{eq:UmovaMasPover}), the action of these operators on a Fock's state describing creation and annihilation processes of particles of mass $\sqrt{(2m)^{2}+\mathbf{q}_{1}^{2} }$, or of the smallest eigenvalue of operator $\hat{H}^{\rm{internal}}$, once interaction between particles has been taken into account. This particle can be considered as a particle that consists of two particles with mass ${m}$, similar to how the two-particle scalar field is constructed as a product of two single-particle fields.

Of course, the example examined in this section is purely a model. In the next sections, we consider more interesting cases for experiments.

\section {Pseudo-scalar meson fields constructed from two bi-spinor fields}
\label {Pseuscalar meson fields constructed from two bi-spinor fields}

The aim of this section is to construct the pseudo-scalar field with the help of two bi-spinor fields, keeping in mind that mesons are bound states of quark and antiquarks. Creation and annihilation operators of this field will correspond to creation and annihilation of mesons. Due to internal coordinates such a meson can interact with the gluon field, which will help to describe the production of such a meson in hadron-hadron scattering.

The tensor product of the two bi-spinor fields $\Psi_{s_{2}}(x_{2})$ and $\bar{\Psi}_{s_{1}}(x_{1})$, that correspond to quarks, can be represented as matrix:
\begin{equation}
\begin{split}
{{\Psi }_{{{s}_{1}}{{s}_{2}}}}\left( {{x}_{1}},{{x}_{2}} \right)={{\bar{\Psi }}_{{{s}_{1}}}}\left( {{x}_{1}} \right){{\Psi }_{{{s}_{2}}}}\left( {{x}_{2}} \right)
\end{split}
\label{eq:TenzoDobSpinor}
\end{equation}
This matrix can be decomposed on the basis of 16 matrices $\hat{\Gamma}_{a}, a=1,2,..,16$ \cite{Bogolyubov:1980}, the basis of algebra generated by the Dirac matrices:
\begin{equation}
\begin{split}
{{\Psi }_{{{s}_{1}}{{s}_{2}}}}\left( {{x}_{1}},{{x}_{2}} \right)=\sum\limits_{a=1}^{16}{{{\phi }_{a}}\left( {{x}_{1}},{{x}_{2}} \right){{\left( {{\Gamma }_{a}} \right)}_{{{s}_{1}}{{s}_{2}}}}}.  
\end{split}
\label{eq:po_Gamma}
\end{equation} 
The set of matrices $\hat{\Gamma}_{a}, a=1,2,..,16$ can be decomposed to a subset; each of these subsets create a basis of the invariant subspace with respect to the Lorentz transformation. Accordingly, the functions $\phi_{a}(x_{1},x_{2})$ decomposed into sets of functions, which are transformed according to the irreducible representations of the Lorentz group. Especially, we are interested in the pseudo-scalar representation, which is realized in the one-dimensional subspace spanned by the matrix $\hat{\gamma}^{5}$:
\begin{equation}
\begin{split}
{{\Psi }_{{{s}_{1}}{{s}_{2}}}}\left( {{x}_{1}},{{x}_{2}} \right)=\phi \left( {{x}_{1}},{{x}_{2}} \right)\gamma _{{{s}_{1}}{{s}_{2}}}^{5}+\cdots . 
\end{split}
\label{eq:Gamma5}
\end{equation}
Here the ellipses denote the rest of the terms in the expression Eq.(\ref{eq:po_Gamma}). By taking into account the properties of the matrix $\hat{\Gamma}_{a}$, the term $\phi_{a}(x_{1},x_{2})$ in this expression can be written as:
\begin{equation}
\begin{split}
\phi \left( {{x}_{1}},{{x}_{2}} \right)=\frac{1}{4}\left( {{\Psi }_{{{s}_{1}}{{s}_{2}}}}\left( {{x}_{1}},{{x}_{2}} \right)\gamma _{{{s}_{1}}{{s}_{2}}}^{5} \right),
\end{split}
\label{eq:koef_pri_gamma5}
\end{equation}
where there is a sum over repeated indices $s_{1}$ and $s_{2}$.

In this section we would like to find the dynamical equation which defines the two-particle function $\phi_{a}(x_{1},x_{2})$. 

Thus we require that the dynamical operator that will be included in this equation should conserve the structure of the invariant subspaces with respect to the Lorentz group that is defined by Eq.(\ref{eq:po_Gamma}). Hence as a result of the action of this operator, the element of each of the invariant subspaces should be mapped to the elements of the same subspace. In other words, the subspaces that are invariant under Lorentz transformations, should remain invariant with respect to dynamical operator of the two-particle field $\Psi_{s_{1}s_{2}}(x_{1}, x_{2})$. 
From this point of view, one cannot apply the same method for a two-particle bi-spinor field that was used for the scalar field in the previous section. Indeed if one will start from the Dirac system of equations by analogy used in the previous section:
\begin{equation}
\begin{cases} 
\scalebox{1.2}{$ i\frac{\partial {{{\bar{\Psi }}}_{{{s}_{1}}}}\left( {{x}_{1}} \right)}{\partial x_{1}^{{{a}_{1}}}}\gamma _{{{s}_{1}}{{s}_{2}}}^{{{a}_{1}}}+m{{{\bar{\Psi }}}_{{{s}_{2}}}}\left( {{x}_{1}} \right)=0, $}\\ 
\scalebox{1.2}{$ i\gamma _{{{s}_{1}}{{s}_{2}}}^{{{a}_{2}}}\frac{\partial {{\Psi }_{{{s}_{2}}}}\left( {{x}_{2}} \right)}{\partial x_{2}^{{{a}_{2}}}}-m{{\Psi }_{{{s}_{1}}}}\left( {{x}_{2}} \right)=0, $}\\ 
\end{cases}
\label{eq:dvaDiraca}
\end{equation} 
by multiplying the first equation by $\Psi_{s_{3}}(x_{2})$, and the second one by $\bar{\Psi}_{s_{3}}(x_{1})$ for two-particle field Eq.(\ref{eq:TenzoDobSpinor}), we obtain the system:
\begin{equation}
\begin{cases} 
\scalebox{1.2}{$i\frac{\partial \Psi_{s_{1}s_{3}}\left( x_{1},x_{2} \right) } {\partial x_{1}^{a_{1}} }\gamma _{s_{1}s_{2}}^{a_{1}}+m{\partial \Psi_ {s_{2}s_{3}} } \left( x_{1},x_{2} \right)=0, $}\\ 
\scalebox{1.2}{$i\gamma_{s_{1}s_{2}}^{a_{2}}\frac{\partial \Psi_{s_{3}s_{2}}\left( x_{1},x_{2} \right) } {\partial x_{2}^{a_{2}} }-m{\Psi_{s_{3}s_{1}} }  \left( x_{1},x_{2} \right)=0. $} \\ 
\end{cases}
\label{eq:dvaDiraca1}
\end{equation}
However substituting into Eq.(\ref{eq:dvaDiraca1}) the expansion Eq.(\ref{eq:po_Gamma}) one may note, that due to coupling of the Dirac matrix with operators on the left hand side of Eq.(\ref{eq:dvaDiraca1}) they may transfer an element from one of the invariant subspaces with respect to the Lorentz group to another invariant subspace by acting on it.

In order to overcome this issue one may use the fact that the components of the bi-spinor should also satisfy the Klein-Gordon-Fock equation in addition to Dirac equation:
\begin{equation}
\begin{cases} 
\scalebox{1.2}{$ -g^{a_{1}a_{2}}\frac{\partial^{2} {{{\bar{\Psi }}}_{{{s}_{1}}}}\left( {{x}_{1}} \right)}{\partial x_{1}^{{{a}_{1}}}\partial x_{1}^{{{a}_{2}}}}-{{m}^{2}}{{{\bar{\Psi }}}_{{{s}_{1}}}}\left( {{x}_{1}} \right)=0, $}\\ 
\scalebox{1.2}{$ -g^{a_{1}a_{2}}\frac{\partial^{2} {{\Psi }_{{{s}_{2}}}}\left( {{x}_{2}} \right)}{\partial x_{2}^{{{a}_{1}}}\partial x_{2}^{{{a}_{2}}}}-{{m}^{2}}{{\Psi }_{{{s}_{2}}}}\left( {{x}_{2}} \right)=0. $} \\ 
\end{cases}
\label{eq:KGF_Dirac}
\end{equation}
By considering a linear combination of Eqs.(\ref{eq:KGF_Dirac}) with coefficients $\Psi_{s_{2}}(x_{2})$ and $\bar{\Psi}_{s_{1}}(x_{1})$ in the same way as in the previous section, we get:
\begin{equation}
\begin{split}
-g^{a_{1}a_{2}}\frac{\partial^{2} {{\Psi }_{{{s}_{1}}{{s}_{2}}}}\left( {{x}_{1}},{{x}_{2}} \right)}{\partial x_{1}^{{{a}_{1}}}\partial x_{1}^{{{a}_{2}}}}
&-g^{a_{1}a_{2}}\frac{\partial^{2} {{\Psi }_{{{s}_{1}}{{s}_{2}}}}\left( {{x}_{1}},{{x}_{2}} \right)}{\partial x_{2}^{{{a}_{1}}}\partial x_{2}^{{{a}_{2}}}}\\ 
& -2{{m}^{2}}{{\Psi }_{{{s}_{1}}{{s}_{2}}}}\left( {{x}_{1}},{{x}_{2}} \right)=0. \\ 
\end{split}
\label{eq:KGF_Dirac1}
\end{equation}
The dynamical operator on the left hand side now conserves the invariant subspaces in Eq.(\ref{eq:po_Gamma}). Therefore, by convoluting Eq.(\ref{eq:KGF_Dirac1}) with the matrix $\hat{\gamma}^{5}$, one will obtain equations for two-particle pseudo-scalar field $\phi(x_{1}, x_{2})$ Eqs.(\ref{eq:Gamma5}),(\ref{eq:koef_pri_gamma5}):
\begin{equation}
\begin{split}
-g^{a_{1}a_{2}}\frac{\partial^{2} \phi \left( {{x}_{1}},{{x}_{2}} \right)}{\partial x_{1}^{{{a}_{1}}}\partial x_{1}^{{{a}_{2}}}}-g^{a_{1}a_{2}}\frac{\partial^{2} \phi \left( {{x}_{1}},{{x}_{2}} \right)}{\partial x_{2}^{{{a}_{1}}}\partial x_{2}^{{{a}_{2}}}} - 2{{m}^{2}}\phi \left( {{x}_{1}},{{x}_{2}} \right)=0. \\ 
\end{split}
\label{eq:KGFfi}
\end{equation}

Note, that from the definitions of Eqs.(\ref{eq:TenzoDobSpinor})-(\ref{eq:koef_pri_gamma5}) it follows that $\phi(x_{1}, x_{2})$ is complex. Hence, instead of solving Eq.(\ref{eq:KGFfi}) 
for real and imaginary parts of this function let's add the complex conjugated terms to it and consider the following system:
\begin{equation}
\begin{cases} 
-g^{a_{1}a_{2}}\frac{\partial^{2} \phi \left( {{x}_{1}},{{x}_{2}} \right)}{\partial x_{1}^{{{a}_{1}}}\partial x_{1}^{{{a}_{2}}}} -g^{a_{1}a_{2}}\frac{\partial^{2} \phi \left( {{x}_{1}},{{x}_{2}} \right)}{\partial x_{2}^{{{a}_{1}}}\partial x_{2}^{{{a}_{2}}}} - 2{{m}^{2}}\phi \left( {{x}_{1}},{{x}_{2}} \right)=0, \\ 
-g^{a_{1}a_{2}}\frac{\partial^{2} {{\phi }^{*}}\left( {{x}_{1}},{{x}_{2}} \right)}{\partial x_{1}^{{{a}_{1}}}\partial x_{1}^{{{a}_{2}}}} -g^{a_{1}a_{2}}\frac{\partial^{2} {{\phi }^{*}}\left( {{x}_{1}},{{x}_{2}} \right)}{\partial x_{2}^{{{a}_{1}}}\partial x_{2}^{{{a}_{2}}}} - 2{{m}^{2}}{{\phi }^{*}}\left( {{x}_{1}},{{x}_{2}} \right)=0. \\ 
\end{cases}
\label{eq:KGFfi111}
\end{equation}
This is the system of Euler-Lagrange equations for Lagrangian:
\begin{equation}
\begin{split}
L\left( {{x}_{1}},{{x}_{2}} \right)&={{g}^{{{a}_{1}}{{a}_{2}}}}\frac{\partial {{\phi }^{*}}\left( {{x}_{1}},{{x}_{2}} \right)}{\partial x_{1}^{{{a}_{1}}}}\frac{\partial \phi \left( {{x}_{1}},{{x}_{2}} \right)}{\partial x_{1}^{{{a}_{2}}}} \\ 
& +{{g}^{{{a}_{1}}{{a}_{2}}}}\frac{\partial {{\phi }^{*}}\left( {{x}_{1}},{{x}_{2}} \right)}{\partial x_{2}^{{{a}_{1}}}}\frac{\partial \phi \left( {{x}_{1}},{{x}_{2}} \right)}{\partial x_{2}^{{{a}_{2}}}} \\ 
& -2{{m}^{2}}{{\phi }^{*}}\left( {{x}_{1}},{{x}_{2}} \right)\phi \left( {{x}_{1}},{{x}_{2}} \right). \\ 
\end{split}
\label{eq:psevdoLag}
\end{equation}

As bi-spinor fields we consider those fields that correspond to quarks and antiquarks. In addition to bi-spinor indices these fields have color indices, denoted as $c_{1}$ and $c_{2}$, and flavor indices, denoted as $f_{1}$ and $f_{2}$. Then the two-particle field will transform, not only with respect to the tensor product of the Lorentz representations, but also with respect to the tensor product of the $SU_{c}(3)$ and $SU_{f}(3)$ group representations. This two-particle field will be denoted as:
\begin{equation}
{{\Psi }_{{{s}_{1}}{{s}_{2}};{{c}_{1}}{{c}_{2}};{{f}_{1}}{{f}_{2}}}}\left( {{x}_{1}},{{x}_{2}} \right)= {{{\bar{\Psi }}}_{{{s}_{1}}{{c}_{1}}{{f}_{1}}}}\left( {{x}_{1}} \right){{\Psi }_{{{s}_{2}}{{c}_{2}}{{f}_{2}}}}\left( {{x}_{2}} \right). \\ 
\label{eq:z_kolorom_i_flavorom}
\end{equation}
 
Instead of Eq.(\ref{eq:koef_pri_gamma5}) we write:
\begin{equation}
\begin{split}
\phi_{c_{1}c_{2}}^{f_{1}f_{2}}\left( {{x}_{1}},{{x}_{2}} \right)=\frac{1}{4}{{{\bar{\Psi }}}_{{{s}_{1}}{{c}_{1}}{{f}_{1}}}}\left( {{x}_{1}} \right){{\Psi }_{{{s}_{2}}{{c}_{2}}{{f}_{2}}}}\left( {{x}_{2}} \right)\gamma _{{{s}_{1}}{{s}_{2}}}^{5}. \\ 
\end{split}
\label{eq:svertca_s_gmma5}
\end{equation}
Here, the flavor indices are written as superscripts just in order to squeeze the notation. Respectively, instead of Lagrangian Eq.(\ref{eq:psevdoLag}) we get:
\begin{equation}
\begin{split}
L&={{g}^{{{a}_{1}}{{a}_{2}}}}\frac{\partial {{\left( \phi _{{{c}_{1}}{{c}_{2}}}^{{{f}_{1}}{{f}_{2}}}\left( {{x}_{1}},{{x}_{2}} \right) \right)}^{*}}}{\partial x_{1}^{{{a}_{1}}}}\frac{\partial \phi _{{{c}_{1}}{{c}_{2}}}^{{{f}_{1}}{{f}_{2}}}\left( {{x}_{1}},{{x}_{2}} \right)}{\partial x_{1}^{{{a}_{2}}}} \\ 
& +{{g}^{{{a}_{1}}{{a}_{2}}}}\frac{\partial {{\left( \phi _{{{c}_{1}}{{c}_{2}}}^{{{f}_{1}}{{f}_{2}}}\left( {{x}_{1}},{{x}_{2}} \right) \right)}^{*}}}{\partial x_{2}^{{{a}_{1}}}}\frac{\partial \phi _{{{c}_{1}}{{c}_{2}}}^{{{f}_{1}}{{f}_{2}}}\left( {{x}_{1}},{{x}_{2}} \right)}{\partial x_{2}^{{{a}_{2}}}} \\ 
& -2{{m}^{2}}{{\left( \phi _{{{c}_{1}}{{c}_{2}}}^{{{f}_{1}}{{f}_{2}}}\left( {{x}_{1}},{{x}_{2}} \right) \right)}^{*}}\phi _{{{c}_{1}}{{c}_{2}}}^{{{f}_{1}}{{f}_{2}}}\left( {{x}_{1}},{{x}_{2}} \right). \\ 
\end{split}
\label{eq:psevdoLag1}
\end{equation}

Now one can define the interaction of the two-particle pseudo-scalar field $\phi_{c_{1}c_{2}}^{f_{1}f_{2}} (x_{1},x_{2})$ with the gluon field by the extension of the derivatives. 
This will ensure that local $SU_{c}(3)$ invariance of the two-particle pseudo-scalar Lagrangian is satisfied, and we can write:
\begin{equation}
\begin{split}
& L=\\
& {{g}^{{{a}_{1}}{{a}_{2}}}} \left( \frac{\partial {{\left( \phi _{{{c}_{1}}{{c}_{2}}}^{{{f}_{1}}{{f}_{2}}}\left( {{x}_{1}},{{x}_{2}} \right) \right)}^{*}}}{\partial x_{1}^{{{a}_{1}}}}  +igA_{{{a}_{1}}}^{{{g}_{1}}}\left( {{x}_{1}} \right){{\left( \phi _{{{c}_{3}}{{c}_{2}}}^{{{f}_{1}}{{f}_{2}}}\left( {{x}_{1}},{{x}_{2}} \right) \right)}^{*}}{{\left( \lambda _{{{c}_{3}}{{c}_{1}}}^{{{g}_{1}}} \right)}^{*}} \right)  \\ 
& \times \left( \frac{\partial \phi _{{{c}_{1}}{{c}_{2}}}^{{{f}_{1}}{{f}_{2}}}\left( {{x}_{1}},{{x}_{2}} \right)}{\partial x_{1}^{{{a}_{2}}}} -igA_{{{a}_{2}}}^{{{g}_{2}}}\left( {{x}_{1}} \right)\lambda _{{{c}_{2}}{{c}_{4}}}^{{{g}_{2}}}\phi _{{{c}_{1}}{{c}_{4}}}^{{{f}_{1}}{{f}_{2}}}\left( {{x}_{1}},{{x}_{2}} \right) \right) \\ 
& +{{g}^{{{a}_{1}}{{a}_{2}}}}\left( \frac{\partial {{\left( \phi _{{{c}_{1}}{{c}_{2}}}^{{{f}_{1}}{{f}_{2}}}\left( {{x}_{1}},{{x}_{2}} \right) \right)}^{*}}}{\partial x_{2}^{{{a}_{1}}}} +igA_{{{a}_{1}}}^{{{g}_{1}}}\left( {{x}_{2}} \right){{\left( \phi _{{{c}_{3}}{{c}_{2}}}^{{{f}_{1}}{{f}_{2}}}\left( {{x}_{1}},{{x}_{2}} \right) \right)}^{*}}{{\left( \lambda _{{{c}_{3}}{{c}_{1}}}^{{{g}_{1}}} \right)}^{*}} \right)  \\ 
& \times \left( \frac{\partial \phi _{{{c}_{1}}{{c}_{2}}}^{{{f}_{1}}{{f}_{2}}}\left( {{x}_{1}},{{x}_{2}} \right)}{\partial x_{2}^{{{a}_{2}}}} -igA_{{{a}_{2}}}^{{{g}_{2}}}\left( {{x}_{2}} \right)\lambda _{{{c}_{2}}{{c}_{4}}}^{{{g}_{2}}}\phi _{{{c}_{1}}{{c}_{4}}}^{{{f}_{1}}{{f}_{2}}}\left( {{x}_{1}},{{x}_{2}} \right) \right) \\ 
& -2{{m}^{2}}{{\left( \phi _{{{c}_{1}}{{c}_{2}}}^{{{f}_{1}}{{f}_{2}}}\left( {{x}_{1}},{{x}_{2}} \right) \right)}^{*}}\phi _{{{c}_{1}}{{c}_{2}}}^{{{f}_{1}}{{f}_{2}}}\left( {{x}_{1}},{{x}_{2}} \right). \\ 
\end{split}
\label{eq:psevdoLag_A}
\end{equation}
Here $g$ is the strong interaction coupling constant, $g_{1}$ and $g_{2}$ are the internal indices of the gluon field.

The transformations rule of the two-particle field $\phi_{c_{1},c_{2}}^{f_{1},f_{2}} (x_{1},x_{2})$ under $SU_{c}(3)$ is coming from Eq.(\ref{eq:z_kolorom_i_flavorom}), and can be written in the form:
\begin{equation}
{\phi' }_{{{c}_{1}}{{c}_{2}}}^{{{f}_{1}}{{f}_{2}}}\left( {{x}_{1}},{{x}_{2}} \right)={{U}_{{{c}_{2}}{{c}_{4}}}}\left( {{x}_{2}} \right)\phi _{{{c}_{3}}{{c}_{4}}}^{{{f}_{1}}{{f}_{2}}}\left( {{x}_{1}},{{x}_{2}} \right)U_{{{c}_{3}}{{c}_{1}}}^{{\dagger }}\left( {{x}_{1}} \right).
\label{eq:SU3c}
\end{equation}
Here $U_{c_{2},c_{4}}(x_{1})$ and $U_{c_{3},c_{1}}(x_{2})^{\dagger}$ are the elements of the coordinate-dependent $SU_{c}(3)$ matrix and its Hermitian conjugated matrix, respectively.

Let us examine, in the special case, the representations of the global transformation group of $SU_{c}(3)$ on the linear space of two index color tensors using the transformation rule Eq.(\ref{eq:SU3c}), one may select the one-dimensional invariant subspace on trivial representation, 
\begin{equation}
\begin{split}
& {{\left( \phi _{{{c}_{1}}{{c}_{2}}}^{{{f}_{1}}{{f}_{2}}}\left( {{x}_{1}},{{x}_{2}} \right) \right)}^{*}}={{\delta }_{{{c}_{1}}{{c}_{2}}}}\phi _{{{f}_{1}}{{f}_{2}}}^{*}\left( {{x}_{1}},{{x}_{2}} \right), \\ 
& \phi _{{{c}_{1}}{{c}_{2}}}^{{{f}_{1}}{{f}_{2}}}\left( {{x}_{1}},{{x}_{2}} \right)={{\delta }_{{{c}_{1}}{{c}_{2}}}}{{\phi }_{{{f}_{1}}{{f}_{2}}}}\left( {{x}_{1}},{{x}_{2}} \right). \\ 
\end{split}
\label{eq:bez_cvet}
\end{equation}

As seen from Eq.(\ref{eq:SU3c}) in the general case of the local transformation (colorless) of a meson could be defined through the manifold of configurations of type:
\begin{equation}
\begin{split}
\phi _{{{c}_{1}}{{c}_{2}}}^{{{f}_{1}}{{f}_{2}}}\left( {{x}_{1}},{{x}_{2}} \right) = \left( {{U}_{{{c}_{2}}{{c}_{3}}}}\left( {{x}_{1}} \right)U_{{{c}_{3}}{{c}_{1}}}^{{\dagger }}\left( {{x}_{2}} \right) \right){{\phi }_{{{f}_{1}}{{f}_{2}}}}\left( {{x}_{1}},{{x}_{2}} \right), \\ 
\end{split}
\label{eq:bezcolor}
\end{equation}
which is gauge equivalent to Eq.(\ref{eq:bez_cvet}).

However, as seen from Eq.(\ref{eq:bezcolor}) the local transformation at $x_{1} \neq x_{2}$ does not conserve the structure of the invariant subspaces with respect to the global transformations. This in turn leads to violation of the local invariance if one will substitute Eq.(\ref{eq:bez_cvet}) into Lagrangian Eq.(\ref{eq:psevdoLag_A}). In order to overcome this issue, let's agree on the following rule. If one will do the local transformation, and then allocate the invariant subspace this will not violate local invariance. But not vice versa. Since these two operations are not permutable. In order to avoid the violation problem we adopt this rule for the multi-particle fields. Firstly we will apply the local transformation for the whole tensor $\phi_{c_{1}c_{2}}^{f_{1}f_{2}}(x_{1},x_{2})$ then we divide it into tensors that are invariant with respect to the global transformations. 

By using the above mentioned rule and substitute Eq.(\ref{eq:bez_cvet}) into Lagrangian Eq.(\ref{eq:psevdoLag_A}) we get:
\begin{equation}
\begin{split}
L & ={{g}^{{{a}_{1}}{{a}_{2}}}}\left( \frac{\partial \phi _{{{f}_{1}}{{f}_{2}}}^{*}\left( {{x}_{1}},{{x}_{2}} \right)}{\partial x_{1}^{{{a}_{1}}}}\frac{\partial {{\phi }_{{{f}_{1}}{{f}_{2}}}}\left( {{x}_{1}},{{x}_{2}} \right)}{\partial x_{1}^{{{a}_{2}}}} \right. \\ 
& \left. +\frac{\partial \phi _{{{f}_{1}}{{f}_{2}}}^{*}\left( {{x}_{1}},{{x}_{2}} \right)}{\partial x_{2}^{{{a}_{1}}}}\frac{\partial {{\phi }_{{{f}_{1}}{{f}_{2}}}}\left( {{x}_{1}},{{x}_{2}} \right)}{\partial x_{2}^{{{a}_{2}}}} \right) \\ 
& -2{{m}^{2}}\phi _{{{f}_{1}}{{f}_{2}}}^{*}\left( {{x}_{1}},{{x}_{2}} \right){{\phi }_{{{f}_{1}}{{f}_{2}}}}\left( {{x}_{1}},{{x}_{2}} \right) \\ 
& +\frac{2}{3}{{g}^{2}}\phi _{{{f}_{1}}{{f}_{2}}}^{*}\left( {{x}_{1}},{{x}_{2}} \right){{\phi }_{{{f}_{1}}{{f}_{2}}}}\left( {{x}_{1}},{{x}_{2}} \right)  \\ 
& \times {{g}^{{{a}_{1}}{{a}_{2}}}}\left( A_{{{a}_{1}}}^{{{g}_{1}}}\left( {{x}_{1}} \right)A_{{{a}_{2}}}^{{{g}_{1}}}\left( {{x}_{1}} \right)+A_{{{a}_{1}}}^{{{g}_{1}}}\left( {{x}_{2}} \right)A_{{{a}_{2}}}^{{{g}_{1}}}\left( {{x}_{2}} \right) \right). \\ 
\end{split}
\label{eq:L_mezon}
\end{equation}
This Lagrangian is locally invariant in the sense discussed above. That is, if we want to make the local transformation we have to return to the previous Lagrangian Eq.(\ref{eq:psevdoLag_A}) and perform the local transformation on it, and then select the invariant subspace Eq.(\ref{eq:bez_cvet}) which is spanned by the unit tensor of color indices.

Applying transformation to the four dimensional Jacobi coordinates in Eq.(\ref{eq:Jacobi}) we again obtain Lagrangian up to a small constant factor on the subset of Eq.(\ref{eq:Odnochasnist}):
\begin{equation}
\begin{split}
L & ={{\left. \left( {{L}_{0}}+{{L}_{{int}}} \right) \right|}_{y_{1}^{0}=0}}, \\ 
L_{0} & ={{g}^{{{a}_{1}}{{a}_{2}}}}\frac{\partial \phi _{{{f}_{1}}{{f}_{2}}}^{*}\left( X,{{y}_{1}} \right)}{\partial {{X}^{{{a}_{1}}}}}\frac{\partial {{\phi }_{{{f}_{1}}{{f}_{2}}}}\left( {{x}_{1}},{{x}_{2}} \right)}{\partial {{X}^{{{a}_{2}}}}}\\ 
& -\left( {{\left( 2m \right)}^{2}}\phi _{{{f}_{1}}{{f}_{2}}}^{*}\left( X,{{y}_{1}} \right){{\phi }_{{{f}_{1}}{{f}_{2}}}}\left( X,{{y}_{1}} \right) \right. \\ 
& \left. + 2\left( 2m \right)\frac{1}{m}\sum\limits_{b=1}^{3}{\frac{\partial \phi _{{{f}_{1}}{{f}_{2}}}^{*}\left( X,{{y}_{1}} \right)}{\partial y_{1}^{b}}\frac{\partial {{\phi }_{{{f}_{1}}{{f}_{2}}}}\left( X,{{y}_{1}} \right)}{\partial y_{1}^{b}}} \right), \\ 
L_{\rm{int}}& =\frac{4}{3}{{g}^{2}}\phi _{{{f}_{1}}{{f}_{2}}}^{*}\left( X,{{y}_{1}} \right){{\phi }_{{{f}_{1}}{{f}_{2}}}}\left( X,{{y}_{1}} \right)  \\ 
& \times {{g}^{{{a}_{1}}{{a}_{2}}}}\left( A_{{{a}_{1}}}^{{{g}_{1}}}\left( X-\frac{1}{2}{{y}_{1}} \right)A_{{{a}_{2}}}^{{{g}_{1}}}\left( X-\frac{1}{2}{{y}_{1}} \right) \right. \\ 
& \left. +A_{{{a}_{1}}}^{{{g}_{1}}}\left( X+\frac{1}{2}{{y}_{1}} \right)A_{{{a}_{2}}}^{{{g}_{1}}}\left( X+\frac{1}{2}{{y}_{1}} \right) \right). \\ 
\end{split}
\label{eq:L0_plus_Lint}
\end{equation}
The Lagrangian $L_{0}$ is similar to the Lagrangian in the previous section and can be considered as the Lagrangian of a free meson field. Which after the quantization procedure correspond to the meson's creation and annihilation operators. Respectively, $L_{\rm{int}}$ is the Lagrangian of interaction for the meson fields with the gauge field. This Lagrangian can be used to describe multi-meson production in scattering processes. Taking into account that in experiments these processes are mostly observed in proton-proton(antiproton) scattering we need to construct the three-particle field that will correspond to protons.

In addition, the criteria of simultaneity in Eq.(\ref{eq:L0_plus_Lint}) leads to the fact that gauge field which interact with the meson field as well as the proton field (this will be shown later in the paper) should be considered as multi-particle fields.

\section{Three-particle bi-spinor field}
\label{Three-particle bi-spinor field}

The three-particle field that will correspond to baryons is obtained by considering all possible products of three bi-spinor field components: 
\begin{equation}
\begin{split}
\Psi _{{{s}_{1}}{{s}_{2}}{{s}_{3}}{{c}_{1}}{{c}_{2}}{{c}_{3}}}^{{{f}_{1}}{{f}_{2}}{{f}_{3}}} & \left( {{x}_{1}},{{x}_{2}},{{x}_{3}} \right) = \\ 
& {{\Psi }_{{{s}_{1}}{{c}_{1}}{{f}_{1}}}}\left( {{x}_{1}} \right){{\Psi }_{{{s}_{2}}{{c}_{2}}{{f}_{2}}}}\left( {{x}_{2}} \right){{\Psi }_{{{s}_{3}}{{c}_{3}}{{f}_{3}}}}\left( {{x}_{3}} \right). \\ 
\end{split}
\label{eq:tri_bispinora}
\end{equation}
For brevity, we will temporarily omit color and flavor indices and consider $\Psi_{s_{1}s_{2}s_{3}}(x_{1}, x_{2},x_{3})$, which under Lorentz transformations transform with respect to the tensor product of the three bi-spinor representations:
\begin{equation}
\begin{split}
{{{{\Psi }'}}_{{{s}_{1}}{{s}_{2}}{{s}_{3}}}} & \left( {{{{x}'}}_{1}},{{{{x}'}}_{2}},{{x'}_{3}} \right)= \\ 
& {{D}_{{{s}_{1}}{{s}_{4}}}}\left( \Lambda  \right){{D}_{{{s}_{2}}{{s}_{5}}}}\left( \Lambda  \right){{D}_{{{s}_{3}}{{s}_{6}}}}\left( \Lambda  \right)  \\ 
& \times {{\Psi }_{{{s}_{4}}{{s}_{5}}{{s}_{6}}}}\left( {{x}_{a}}={{\Lambda }^{-1}}{{{{x}'}}_{a}} \right), ~~a=1,2,3. \\ 
\end{split}
\label{eq:tenzornij_dobutok}
\end{equation}
Here $\Lambda$ is the Lorentz transformation, $D_{s_{1} s_{4}}(\Lambda)$ are the matrix elements of the bi-spinor representation of the Lorentz group. 

From the linear space of the triple index tensors $\Psi_{s_{1}s_{2}s_{3}}(x_{1}, x_{2},x_{3})$ we have to pick out the invariant subspace that transforms with respect to the bi-spinor representation of the Lorentz group. This can be done in different ways.

For instance, one may pick from each tensor $\Psi_{s_{1}s_{2}s_{3}}(x_{1}, x_{2},x_{3})$, the completely antisymmetric part with respect to the three bi-spinor indices. Completely antisymmetric tensors, which will be denoted as $\Psi_{s_{1}s_{2}s_{3}}^{a}(x_{1}, x_{2},x_{3})$ create a four-space invariant subspace on space that corresponds to three index tensors from Eq.(\ref{eq:tenzornij_dobutok}). Convolving each such tensor with Levi-Civita symbol $\varepsilon_{s_{1}s_{2}s_{3}s_{4}}$, we get:
\begin{equation}
\begin{split}
& {{{\bar{\Psi }}}_{{{s}_{1}}}}\left( {{x}_{1}},{{x}_{2}},{{x}_{3}} \right) = {{\varepsilon }_{{{s}_{1}}{{s}_{2}}{{s}_{3}}{{s}_{4}}}}\Psi _{{{s}_{2}}{{s}_{3}}{{s}_{4}}}^{\left( a \right)}\left( {{x}_{1}},{{x}_{2}},{{x}_{3}} \right), \\ 
\end{split}
\label{eq:trexchastichnij_s_chertockoj}
\end{equation}
which transforms with respect to the inverse bi-spinor representation. With help of this quantity, one may construct this expression: 
\begin{equation}
\begin{split}
{{\Psi }_{{{s}_{2}}}}\left( {{x}_{1}},{{x}_{2}},{{x}_{3}} \right)={{\bar{\Psi }}_{{{s}_{1}}}}\left( {{x}_{1}},{{x}_{2}},{{x}_{3}} \right){{\left( {{{\hat{\gamma }}}^{1}}{{{\hat{\gamma }}}^{3}} \right)}_{{{s}_{1}}{{s}_{2}}}},
\end{split}
\label{eq:tri_bisp}
\end{equation}
that transforms with respect to the desired, for us, bi-spinor representation.

Therefore, one may select from the linear subspace corresponding to the tensors of type $\Psi_{s_{1}s_{2}s_{3}}(x_{1}, x_{2},x_{3})$ the subspace of tensors of type:
\begin{equation}
\begin{split}
& {{\Psi }_{{{s}_{1}}{{s}_{2}}{{s}_{3}}}}\left( {{x}_{1}},{{x}_{2}},{{x}_{3}} \right) = - \frac{1}{3!}{{\varepsilon }_{{{s}_{1}}{{s}_{2}}{{s}_{3}}{{s}_{4}}}}{{\Psi }_{{{s}_{5}}}}\left( {{x}_{1}},{{x}_{2}},{{x}_{3}} \right){{\left( {{{\hat{\gamma }}}^{1}}{{{\hat{\gamma }}}^{3}} \right)}_{{{s}_{5}}{{s}_{4}}}}, \\ 
\end{split}
\label{eq:3bispinor}
\end{equation}
where $\Psi_{s_{5}}(x_{1}, x_{2},x_{3})$ is the bi-spinor that is defined by Eq.(\ref{eq:tri_bisp}). On this subspace the Lorentz group representation which is equivalent to the bi-spinor representation can be realized. The subspace, where this representation is realized, can be selected in another way, for instance, by considering bi-spinors from Eq.(\ref{eq:tenzornij_dobutok}) in the chiral representation and expanding each of the bi-spinor space-multipliers into a direct sum of the left and right subspaces. This method does not matter for us, because we are interested primarily in the dynamic equation for the three-particle bi-spinor.

In order to obtain this equation we will proceed with the same reasons as in the previous section. Thus, we start from the Klein-Gordon-Fock equation for the components of each bi-spinor multipliers in Eq.(\ref{eq:tenzornij_dobutok}). With the reasoning that led us to Eq.(\ref{eq:KGF_Dirac1}), we can write:
\begin{equation}
\begin{split}
& -{{g}^{{{a}_{1}}{{a}_{2}}}}\frac{{{\partial }^{2}}{{\Psi }_{{{s}_{1}}{{s}_{2}}{{s}_{3}}}}\left( {{x}_{1}},{{x}_{2}},{{x}_{3}} \right)}{\partial x_{1}^{{{a}_{1}}}\partial x_{1}^{{{a}_{2}}}} - {{g}^{{{a}_{1}}{{a}_{2}}}}\frac{{{\partial }^{2}}{{\Psi }_{{{s}_{1}}{{s}_{2}}{{s}_{3}}}}\left( {{x}_{1}},{{x}_{2}},{{x}_{3}} \right)}{\partial x_{2}^{{{a}_{1}}}\partial x_{2}^{{{a}_{2}}}} \\ 
& -{{g}^{{{a}_{1}}{{a}_{2}}}}\frac{{{\partial }^{2}}{{\Psi }_{{{s}_{1}}{{s}_{2}}{{s}_{3}}}}\left( {{x}_{1}},{{x}_{2}},{{x}_{3}} \right)}{\partial x_{3}^{{{a}_{1}}}\partial x_{3}^{{{a}_{2}}}} -3{{m}^{2}}{{\Psi }_{{{s}_{1}}{{s}_{2}}{{s}_{3}}}}\left( {{x}_{1}},{{x}_{2}},{{x}_{3}} \right)=0. \\ 
\end{split}
\label{eq:KGF3part_spin}
\end{equation}
Here we again simplify our notation by omitting the color and flavor indices.

After transformation of this equation into the Jacobi coordinates of Eq.(\ref{eq:Jacobi}), we get:
\begin{equation}
\begin{split}
&-{{g}^{{{a}_{1}}{{a}_{2}}}}\frac{{{\partial }^{2}}{{\Psi }_{{{s}_{1}}{{s}_{2}}{{s}_{3}}}}\left( X,{{y}_{1}},{{y}_{2}} \right)}{\partial {{X}^{{{a}_{1}}}}\partial {{X}^{{{a}_{2}}}}} -\left( 9{{m}^{2}}{{\Psi }_{{{s}_{1}}{{s}_{2}}{{s}_{3}}}}\left( X,{{y}_{1}},{{y}_{2}} \right) \right. \\ 
&\left. +\frac{9}{2}{{g}^{{{a}_{1}}{{a}_{2}}}}\frac{{{\partial }^{2}}{{\Psi }_{{{s}_{1}}{{s}_{2}}{{s}_{3}}}}\left( X,{{y}_{1}},{{y}_{2}} \right)}{\partial y_{1}^{{{a}_{1}}}\partial y_{1}^{{{a}_{2}}}} + 6{{g}^{{{a}_{1}}{{a}_{2}}}}\frac{{{\partial }^{2}}{{\Psi }_{{{s}_{1}}{{s}_{2}}{{s}_{3}}}}\left( X,{{y}_{1}},{{y}_{2}} \right)}{\partial y_{2}^{{{a}_{1}}}\partial y_{2}^{{{a}_{2}}}} \right)=0 \\ 
\end{split}
\label{eq:Trispin_Jacobi}
\end{equation}
Convolving this equation with the Levi-Civita symbol and then convolving within elements of the matrix $\hat{\gamma}^{1}\hat{\gamma}^{3}$ as this was discussed previously, we obtain the equation for the three bi-spinor Eq.(\ref{eq:tri_bisp}). By examining this equation on the subspace of Eq.(\ref{eq:Odnochasnist}) we have:
\begin{equation}
\begin{split}
& -{{g}^{{{a}_{1}}{{a}_{2}}}}\frac{{{\partial }^{2}}{{\Psi }_{{{s}_{1}}}}\left( X, \mathbf{y}_{1}, \mathbf{y_{2}}  \right)}{\partial {{X}^{{{a}_{1}}}}\partial {{X}^{{{a}_{2}}}}} - \left( {{\left( 3m \right)}^{2}}{{\Psi }_{{{s}_{1}}}}\left( X, \mathbf{y}_{1}, \mathbf{y_{2}} \right) \right. \\ 
& +2\left( 3m \right)\left( -\frac{3}{4m}\Delta_{\mathbf{y}_{1}}{{\Psi }_{{{s}_{1}}}}\left( X, \mathbf{y}_{1}, \mathbf{y_{2}} \right) \right. \\ 
& \left. \left. -\frac{1}{m}\Delta_{\mathbf{y_{2}}}{{\Psi }_{{{s}_{1}}}}\left( X, \mathbf{y}_{1}, \mathbf{y_{2}} \right) \right) \right)=0. \\ 
\end{split}
\label{eq:KGF_3Bisp}
\end{equation}

Again as in Eqs.(\ref{eq:LagEiler})-(\ref{eq:LagEiler1}) we introduce the notation for the three particle system Hamiltonian:
\begin{equation}
\begin{split}
& \hat{H}^{\rm{internal}} = \left( 3m \right)\hat{E}-\frac{3}{4m}\Delta_{\mathbf{y_{1}}} -  \frac{1}{m}\Delta_{\mathbf{y_{2}}} . \\ 
\end{split}
\label{eq:Hinternal3particls}
\end{equation}

Further as in Eq.(\ref{eq:LagEiler1}) with an accuracy of the order of the square of the ratio of the characteristic internal energy of the three-particle system to its rest energy, we can write:
\begin{equation}
\begin{split}
-{{g}^{{{a}_{1}}{{a}_{2}}}}\frac{{{\partial }^{2}}{{\Psi }_{{{s}_{1}}}}\left( X, \mathbf{y}_{1}, \mathbf{y_{2}} \right)}{\partial {{X}^{{{a}_{1}}}}\partial {{X}^{{{a}_{2}}}}} - {{\left( {{{\hat{H}}}^{{\rm{internal}}}} \right)}^{2}}{{\Psi }_{{{s}_{1}}}}\left( X,\mathbf{y}_{1}, \mathbf{y_{2}} \right)=0. \\ 
\end{split}
\label{eq:KGFInternal}
\end{equation}

It is  obvious that the operators $\hat{H}^{\rm{internal}}$ and $i\partial/\partial X^{a}$ for all values of the index ${a}$ commute. This allows us to apply the same factorization procedure 
for the Eq.(\ref{eq:KGFInternal}) as for the ``ordinary"  Klein-Gordon-Fock equation which leads to the ``ordinary" Dirac equation. In our case this factorization gives us the three particle analog of it:
\begin{equation}
\begin{split}
& i\gamma _{{{s}_{1}}{{s}_{2}}}^{a}\frac{\partial {{\Psi }_{{{s}_{2}}}}\left( X, \mathbf{y}_{1}, \mathbf{y_{2}} \right)}{\partial {{X}^{a}}} \\ 
& -\left( 3m-\frac{3}{4m}  \Delta_{\mathbf{y}_{1}}-\frac{1}{m}\Delta_{\mathbf{y_{2}}} \right){{\Psi }_{{{s}_{1}}}}\left( X,\mathbf{y}_{1}, \mathbf{y_{2}} \right)=0. \\ 
\end{split}
\label{eq:Dirac3particls}
\end{equation}

Furthermore, the Eq.(\ref{eq:Dirac3particls}) will be considered  as a dynamical equation for the three particle bi-spinor field, which we are looking for.

For further extension of derivatives in order to describe the interaction with the gluon field, we consider Eq.(\ref{eq:Dirac3particls}) not using the subsets in Eq.(\ref{eq:Odnochasnist}), but by using the full tensor product of three  Minkowski  spaces:
\begin{equation}
\begin{split}
i\gamma _{{{s}_{1}}{{s}_{2}}}^{a}\frac{\partial {{\Psi }_{{{s}_{2}}}}\left( X,{{y}_{1}},{{y}_{2}} \right)}{\partial {{X}^{a}}}  & - \left( 3m+\frac{3}{4m}g^{a_{1}a_{2}}\frac{{{\partial }^{2}}}{\partial y_{1}^{{{a}_{1}}}\partial y_{1}^{{{a}_{2}}}} \right. \\ 
& \left. +  \frac{1}{m}g^{a_{1}a_{2}}\frac{{{\partial }^{2}}}{\partial y_{2}^{{{a}_{1}}}\partial y_{2}^{{{a}_{2}}}} \right){{\Psi }_{{{s}_{1}}}}\left( X,{{y}_{1}},{{y}_{2}} \right)=0. \\ 
\end{split}
\label{eq:Dirac3particls1}
\end{equation}

Now, let us turn back to the initial coordinates:
\begin{equation}
\begin{split}
& i\gamma _{{{s}_{1}}{{s}_{2}}}^{a}\frac{\partial {{\Psi }_{{{s}_{2}}}}\left( {{x}_{1}},{{x}_{2}},{{x}_{3}} \right)}{\partial x_{1}^{a}}+i\gamma _{{{s}_{1}}{{s}_{2}}}^{a}\frac{\partial {{\Psi }_{{{s}_{2}}}}\left( {{x}_{1}},{{x}_{2}},{{x}_{3}} \right)}{\partial x_{2}^{a}} \\ 
& +i\gamma _{{{s}_{1}}{{s}_{2}}}^{a}\frac{\partial {{\Psi }_{{{s}_{2}}}}\left( {{x}_{1}},{{x}_{2}},{{x}_{3}} \right)}{\partial x_{3}^{a}}-3m{{\Psi }_{{{s}_{1}}}}\left( {{x}_{1}},{{x}_{2}},{{x}_{3}} \right) \\ 
& -\frac{1}{3m}\left(  g^{a_{1}a_{3}}\frac{\partial^{2} \Psi_{s_{1}}(x_{1}, x_{2}, x_{3})} {\partial x_{1}^{a_{1}}\partial x_{1}^{a_{3}}} + g^{a_{1}a_{3}}\frac{\partial^{2} \Psi_{s_{1}}(x_{1}, x_{2}, x_{3})} {\partial x_{2}^{a_{1}}\partial x_{2}^{a_{3}}} \right. \\
&+ g^{a_{1}a_{3}}\frac{\partial^{2} \Psi_{s_{1}}(x_{1}, x_{2}, x_{3})} {\partial x_{3}^{a_{1}}\partial x_{3}^{a_{3}}} - g^{a_{1}a_{3}}\frac{\partial^{2} \Psi_{s_{1}}(x_{1}, x_{2}, x_{3})} {\partial x_{1}^{a_{1}}\partial x_{2}^{a_{3}}} \\
& \left. - g^{a_{1}a_{3}}\frac{\partial^{2} \Psi_{s_{1}}(x_{1}, x_{2}, x_{3})} {\partial x_{3}^{a_{1}}\partial x_{1}^{a_{3}}} - g^{a_{1}a_{3}}\frac{\partial^{2} \Psi_{s_{1}}(x_{1}, x_{2}, x_{3})} {\partial x_{3}^{a_{1}}\partial x_{2}^{a_{3}}}\right)=0. \\
\end{split}
\label{eq:Dirac3particlsx1x2x3}
\end{equation}

This equation represents the Euler-Lagrange equation for the Lagrangian:
\begin{equation}
\begin{split}
& L=\frac{i}{2}\left( \bar{\Psi }_{{{s}_{1}}}^{{{c}_{1}}{{c}_{2}}{{c}_{3}}}\gamma _{{{s}_{1}}{{s}_{2}}}^{a}\frac{\partial \Psi _{{{s}_{2}}}^{{{c}_{1}}{{c}_{2}}{{c}_{3}}}}{\partial x_{1}^{a}} - \frac{\partial \bar{\Psi }_{{{s}_{1}}}^{{{c}_{1}}{{c}_{2}}{{c}_{3}}}}{\partial x_{1}^{a}}\gamma _{{{s}_{1}}{{s}_{2}}}^{a}\Psi _{{{s}_{2}}}^{{{c}_{1}}{{c}_{2}}{{c}_{3}}} \right) \\ 
& +\frac{i}{2}\left( \bar{\Psi }_{{{s}_{1}}}^{{{c}_{1}}{{c}_{2}}{{c}_{3}}}\gamma _{{{s}_{1}}{{s}_{2}}}^{a}\frac{\partial \Psi _{{{s}_{2}}}^{{{c}_{1}}{{c}_{2}}{{c}_{3}}}}{\partial x_{2}^{a}} - \frac{\partial \bar{\Psi }_{{{s}_{1}}}^{{{c}_{1}}{{c}_{2}}{{c}_{3}}}}{\partial x_{2}^{a}}\gamma _{{{s}_{1}}{{s}_{2}}}^{a}\Psi _{{{s}_{2}}}^{{{c}_{1}}{{c}_{2}}{{c}_{3}}} \right) \\ 
& +\frac{i}{2}\left( \bar{\Psi }_{{{s}_{1}}}^{{{c}_{1}}{{c}_{2}}{{c}_{3}}}\gamma _{{{s}_{1}}{{s}_{2}}}^{a}\frac{\partial \Psi _{{{s}_{2}}}^{{{c}_{1}}{{c}_{2}}{{c}_{3}}}}{\partial x_{3}^{a}} - \frac{\partial \bar{\Psi }_{{{s}_{1}}}^{{{c}_{1}}{{c}_{2}}{{c}_{3}}}}{\partial x_{3}^{a}}\gamma _{{{s}_{1}}{{s}_{2}}}^{a}\Psi _{{{s}_{2}}}^{{{c}_{1}}{{c}_{2}}{{c}_{3}}} \right)\\
&-3m\bar{\Psi }_{{{s}_{1}}}^{{{c}_{1}}{{c}_{2}}{{c}_{3}}}\Psi _{{{s}_{1}}}^{{{c}_{1}}{{c}_{2}}{{c}_{3}}}\\ 
& +\frac{1}{3m}{{g}^{{{a}_{1}}{{a}_{2}}}}\left( \frac{\partial \bar{\Psi }_{{{s}_{1}}}^{{{c}_{1}}{{c}_{2}}{{c}_{3}}}}{\partial x_{1}^{{{a}_{1}}}}\frac{\partial \Psi _{{{s}_{1}}}^{{{c}_{1}}{{c}_{2}}{{c}_{3}}}}{\partial x_{1}^{{{a}_{2}}}} + \frac{\partial \bar{\Psi }_{{{s}_{1}}}^{{{c}_{1}}{{c}_{2}}{{c}_{3}}}}{\partial x_{2}^{{{a}_{1}}}}\frac{\partial \Psi _{{{s}_{1}}}^{{{c}_{1}}{{c}_{2}}{{c}_{3}}}}{\partial x_{2}^{{{a}_{2}}}}\right. \\
&+\frac{\partial \bar{\Psi }_{{{s}_{1}}}^{{{c}_{1}}{{c}_{2}}{{c}_{3}}}}{\partial x_{3}^{{{a}_{1}}}}\frac{\partial \Psi _{{{s}_{1}}}^{{{c}_{1}}{{c}_{2}}{{c}_{3}}}}{\partial x_{3}^{{{a}_{2}}}} - \frac{\partial \bar{\Psi }_{{{s}_{1}}}^{{{c}_{1}}{{c}_{2}}{{c}_{3}}}}{\partial x_{1}^{{{a}_{1}}}}\frac{\partial \Psi _{{{s}_{1}}}^{{{c}_{1}}{{c}_{2}}{{c}_{3}}}}{\partial x_{2}^{{{a}_{2}}}}\\
& \left. -\frac{\partial \bar{\Psi }_{{{s}_{1}}}^{{{c}_{1}}{{c}_{2}}{{c}_{3}}}}{\partial x_{3}^{{{a}_{1}}}}\frac{\partial \Psi _{{{s}_{1}}}^{{{c}_{1}}{{c}_{2}}{{c}_{3}}}}{\partial x_{1}^{{{a}_{2}}}} - \frac{\partial \bar{\Psi }_{{{s}_{1}}}^{{{c}_{1}}{{c}_{2}}{{c}_{3}}}}{\partial x_{3}^{{{a}_{1}}}}\frac{\partial \Psi _{{{s}_{1}}}^{{{c}_{1}}{{c}_{2}}{{c}_{3}}}}{\partial x_{2}^{{{a}_{2}}}} \right). \\  
\end{split}
\label{eq:Dirac3Lagrangian}
\end{equation}
In order to make the notation less bulky we did not write the arguments $(x_{1}, x_{2},x_{3})$ for the field functions $\Psi_{s_{1};f_{1}f_{2}f_{3}}^{c_{1}c_{2}c_{3}}(x_{1}, x_{2},x_{3})$ and
$\bar{\Psi}_{s_{1};f_{1}f_{2}f_{3}}^{c_{1}c_{2}c_{3}}(x_{1}, x_{2},x_{3})$. Also we temporarily omitted the flavor indices $f_{1}, f_{2}, f_{3}$. Just color indices $c_{1}, c_{2}, c_{3}$ have been written as superscripts.

Now one may extend the derivatives in this Lagrangian in the same way as this was done in case of the meson. The colorlessness of the baryon can be expressed by determining the dependencies of the three-particle field functions with respect to color indices with the help of the Levi-Civita symbol  $\varepsilon_{c_{1}c_{2}c_{3}}$ as:
\begin{equation}
\begin{split}
& \bar{\Psi }_{s_{1}; f_{1}f_{2}f_{3}}^{c_{1}c_{2}c_{3}}\left( {{x}_{1}},{{x}_{2}},{{x}_{3}} \right) ={{\varepsilon }_{{{c}_{1}}{{c}_{2}}{{c}_{3}}}}\bar{\Psi }_{{{s}_{1}}}^{f_{1}f_{2}f_{3}}\left( {{x}_{1}},{{x}_{2}},{{x}_{3}} \right), \\ 
& \Psi _{s_{1}; f_{1}f_{2}f_{3}}^{c_{1}c_{2}c_{3}}\left( {{x}_{1}},{{x}_{2}},{{x}_{3}} \right)={{\varepsilon }_{{{c}_{1}}{{c}_{2}}{{c}_{3}}}}\Psi _{{{s}_{1}}}^{f_{1}f_{2}f_{3}}\left( {{x}_{1}},{{x}_{2}},{{x}_{3}} \right). \\ 
\end{split}
\label{eq:color_Levi_Chivitta}
\end{equation}

Again, as in case of meson fields with local $SU(3)$ transformations, we can write 
\begin{equation}
\begin{split}
& {\bar{\Psi}^{\prime ~c_{1}c_{2}c_{3}}}_{s_{1};f_{1}f_{2}f_{3}}(x_{1}, x_{2},x_{3})=\\
& {\bar{\Psi}^{\prime~c_{4}c_{5}c_{6}}}_{s_{1};f_{1}f_{2}f_{3}}(x_{1}, x_{2},x_{3})U^{\dagger}_{c_{4}c_{1}}(x_{1})U^{\dagger}_{c_{5}c_{2}}(x_{2})U^{\dagger}_{c_{6}c_{3}}(x_{3})\\
& {\Psi}^{\prime ~c_{1}c_{2}c_{3}}_{s_{1};f_{1}f_{2}f_{3}}(x_{1}, x_{2},x_{3})=\\
& U_{c_{1}c_{4}}(x_{1})U_{c_{2}c_{5}}(x_{2})U_{c_{3}c_{6}}(x_{3}){\Psi}^{\prime~c_{4}c_{5}c_{6}}_{s_{1};f_{1}f_{2}f_{3}}(x_{1}, x_{2},x_{3}),\\
\end{split}
\label{eq:localnoe_SU3color3part}
\end{equation}
where we use notation similar to the Eq.(\ref{eq:SU3c}). Here the dependency from color indices like in Eq.(\ref{eq:color_Levi_Chivitta}) are not preserved, and we have an infinite set of the gauge equivalence field configurations as a result of the colorless combinations. Such a case is typical to the single-particle gauge theories, when in addition to each of the field configurations there is a countless number of the gauge equivalent configurations, which are physically indistinguishable from each other in the sense that in the experiment, all these configurations will manifest the same.  But, in the multi-particle theory appears the feature consisting in the fact that by colorless should be considered not only the dependence on the color index, which at the global $SU(3)$ transformation transform into itself, but also any dependence obtained from it with the help of the local gauge transformations.
Local invariance of the Lagrangian can be obtained in the same way as was done in the previous section. Namely, first transform the linear space of tensors and then select the invariant subspace on it. 
By doing this we get the Lagrangian of the three-particle bi-spinor field which interacts with the gluon field as $L=L_{0}+L_{\rm{int}}$, where $L_{0}$ is defined by Eq.(\ref{eq:Dirac3Lagrangian}), $L_{\rm{int}}$ using the color configuration from Eq.(\ref{eq:color_Levi_Chivitta}) has the form:
\begin{equation}
\begin{split}
{{L}_{{\rm{int}}}}&=\frac{1}{9m}{{g}^{2}}{{{\bar{\Psi }}}_{{{s}_{1}}}}\left( {{x}_{1}},{{x}_{2}},{{x}_{3}} \right){{\Psi }_{{{s}_{1}}}}\left( {{x}_{1}},{{x}_{2}},{{x}_{3}} \right) \\ 
& \times \left( \left( \phi \left( {{x}_{1}},{{x}_{2}} \right)+\chi \left( {{x}_{1}},{{x}_{2}} \right) \right) \right. \\ 
& +\left( \phi \left( {{x}_{1}},{{x}_{3}} \right)+\chi \left( {{x}_{1}},{{x}_{3}} \right) \right) \\ 
& \left. +\left( \phi \left( {{x}_{2}},{{x}_{3}} \right)+\chi \left( {{x}_{2}},{{x}_{3}} \right) \right) \right), \\ 
\end{split}
\label{eq:LintProton}
\end{equation}
here we use notations similar to Eq.(\ref{eq:psevdoLag_A}). In addition, the following replacement has been applied:
\begin{equation}
\begin{split}
\phi \left( {{x}_{{{b}_{1}}}},{{x}_{{{b}_{2}}}} \right) & ={{g}^{{{a}_{1}}{{a}_{2}}}}{{A}_{{{a}_{1}},{{g}_{1}}}}\left( {{x}_{{{b}_{1}}}} \right){{A}_{{{a}_{2}},{{g}_{1}}}}\left( {{x}_{{{b}_{2}}}} \right), \\ 
\chi \left( {{x}_{{{b}_{1}}}},{{x}_{{{b}_{2}}}} \right) & ={{g}^{{{a}_{1}}{{a}_{2}}}}\left( {{A}_{{{a}_{1}},{{g}_{1}}}}\left( {{x}_{{{b}_{1}}}} \right){{A}_{{{a}_{2}},{{g}_{1}}}}\left( {{x}_{{{b}_{1}}}} \right) \right.\\ 
& \left. +{{A}_{{{a}_{1}},{{g}_{1}}}}\left( {{x}_{{{b}_{2}}}} \right){{A}_{{{a}_{2}},{{g}_{1}}}}\left( {{x}_{{{b}_{2}}}} \right) \right). \\ 
\end{split}
\label{eq:Poznachenna_fi_xi}
\end{equation}

For the Eq.(\ref{eq:LintProton}) one can choose the 
Jacobi coordinates and examine it on the subset of Eq.(\ref{eq:Odnochasnist}). Therefore, one can describe proton interactions with the gluon field, which can interact with meson fields generating secondary mesons. This allows us to describe processes of inelastic and elastic proton scattering within the framework of multi-particle fields. However as was mentioned before the products of the gluon field functions, which were included into $L_{\rm{int}}$ Eq.(\ref{eq:L0_plus_Lint}) and Eq.(\ref{eq:LintProton}), are also examined on the subset of Eq.(\ref{eq:Odnochasnist}). That is why for the construction of these models we have to consider  also a two-particle gluon field.

\section{Two-particle gluon field}
\label{Two-particle gluon field}

As seen from $L_{\rm{int}}$ in Eqs.(\ref{eq:L0_plus_Lint}),(\ref{eq:LintProton}) the gluon field $A_{a_{1},g_{1}}(x_{1})$ is included in the scalar combinations in Eq.(\ref{eq:Poznachenna_fi_xi}) with respect to the Lorentz transformation, as well as with respect to the adjoint representation of the global $SU_{c}(3)$ group. We will consider Eq.(\ref{eq:Poznachenna_fi_xi}) as two-particle fields. However, examination of the multi-particle gauge field from a physical point of view requires answers to several questions.

The consideration of the multi-particle fields in the previous sections was based on the fact that hadrons contain a certain amount of constituent quarks. In addition, those quarks have such mass that the energy of their interaction is not large enough for creation of new quarks. Therefore, the number of constituent quarks is fixed. This allows one to examine the internal hadron state using a non-relativistic approximation. But there is a question with respect to the gauge field: what it represents from a physical point of view? If one considers the two-particle gauge field, what particles will represent the quanta of this field after the quantization procedure? How are they composed? How will the fact that this field is a two-partial field be manifest? We will try to give answers to these questions in this section, since these answers will be based on properties of solutions of the dynamical equations for two-particle gauge field. Let's consider these equations.

By multiplying each of the Euler-Lagrange equations for non-Abilian gauge field $A_{a_{1},g_{1}}(x_{1})$ with each component for the field $A_{a_{2},g_{2}}(x_{2})$, we have:
\begin{equation}
\begin{split}
& {{A}_{{{a}_{4}},{{g}_{4}}}}\left( {{x}_{2}} \right){{g}^{{{a}_{3}}{{a}_{2}}}}  {{{\hat{D}}}_{{{a}_{3}}}}\left( {{A}_{{{a}_{3}},{{g}_{3}}}}\left( {{x}_{1}} \right) \right){{F}_{{{a}_{1}}{{a}_{2}},{{g}_{1}}}}\left( {{A}_{{{a}_{k}},{{g}_{k}}}}\left( {{x}_{1}} \right) \right)=0. \\ 
\end{split}
\label{eq:Gluon_Lagrang_Eiler_naAx2}
\end{equation}
Here we introduce the following notation for the tensor of the gauge field:
\begin{equation}
\begin{split}
& {{F}_{{{a}_{1}}{{a}_{2}},{{g}_{1}}}}\left( {{A}_{{{a}_{k,}}{{g}_{k}}}}\left( {{x}_{1}} \right) \right) =\\
& \frac{\partial {{A}_{{{a}_{1}},{{g}_{1}}}}\left( {{x}_{1}} \right)}{\partial x_{1}^{{{a}_{2}}}}  -\frac{\partial {{A}_{{{a}_{2}},{{g}_{1}}}}\left( {{x}_{1}} \right)}{\partial x_{1}^{{{a}_{1}}}} - gc_{g_{1}g_{2}g_{3}}{{A}_{{{a}_{1}},{{g}_{2}}}}\left( {{x}_{1}} \right){{A}_{{{a}_{2}},{{g}_{3}}}}\left( {{x}_{1}} \right), \\ 
\end{split}
\label{eq:tenzor_pola}
\end{equation} 
where $c_{g_{1}g_{2}g_{3}}$ are the structure constants of the gauge group. Using the following notation $F_{a_{1}a_{2},g_{1}} \left( A_{a_{k},g_{k}} (x_{1}) \right)$ we would like to emphasize that here we are considering the equation with respect to the field $A_{a_{1},g_{1}} (x_{1})$ and the field tensor just used as notation. Moreover, in Eq.(\ref{eq:Gluon_Lagrang_Eiler_naAx2}) we use the notation for the extended derivative in the adjoint representation of the gauge group:
\begin{equation}
\begin{split}
{{{\hat{D}}}_{{{a}_{3}}}}  \left( {{A}_{{{a}_{3}},{{g}_{3}}}}\left( {{x}_{1}} \right) \right)&{{F}_{{{a}_{1}}{{a}_{2}},{{g}_{1}}}}\left( {{A}_{{{a}_{k}},{{g}_{k}}}}\left( {{x}_{1}} \right) \right)= \\ 
& \frac{\partial {{F}_{{{a}_{1}}{{a}_{2}},{{g}_{1}}}}\left( {{A}_{{{a}_{k}},{{g}_{k}}}}\left( {{x}_{1}} \right) \right)}{\partial x_{1}^{{{a}_{3}}}}\\
- & g{{A}_{{{a}_{3}},{{g}_{3}}}}\left( {{x}_{1}} \right){{c}_{{{g}_{3}}{{g}_{1}}{{g}_{2}}}}{{F}_{{{a}_{1}}{{a}_{2}},{{g}_{2}}}}\left( {{A}_{{{a}_{k}},{{g}_{k}}}}\left( {{x}_{1}} \right) \right). \\ 
\end{split}
\label{eq:kovariantna_poxidna}
\end{equation}

Note, that the system of Eqs.(\ref{eq:Gluon_Lagrang_Eiler_naAx2}) is invariant with respect to local gauge transformation. Indeed, expressing $A_{a_{1},g_{1}} (x_{1})$ and $A_{a_{2},g_{2}} (x_{2})$ fields by gauge equivalent field configurations, we have:
\begin{equation}
\begin{split}
& {{\left( \exp \left( {{{\hat{I}}}_{{{g}_{5}}}}{{\theta }_{{{g}_{5}}}}\left( {{x}_{2}} \right) \right) \right)}_{g_{4}g_{6}} }{{\left( \exp \left( {{{\hat{I}}}_{{{g}_{7}}}}{{\theta }_{{{g}_{7}}}}\left( {{x}_{1}} \right) \right) \right)}_{g_{1}g_{8}} } \\ 
& \times {{{{A}'}}_{{{a}_{4}},{{g}_{6}}}}\left( {{x}_{2}} \right){{g}^{{{a}_{3}}{{a}_{2}}}} {{{\hat{D}}}_{{{a}_{3}}}}\left( {{{{A}'}}_{{{a}_{3}},{{g}_{3}}}}\left( {{x}_{1}} \right) \right){{F}_{{{a}_{1}}{{a}_{2}},{{g}_{8}}}}\left( {{{{A}'}}_{{{a}_{k}},{{g}_{k}}}}\left( {{x}_{1}} \right) \right) \\ 
& +\frac{\partial {{\theta }_{{{g}_{4}}}}\left( {{x}_{2}} \right)}{\partial {{x}^{a}}}{{\left( \exp \left( {{{\hat{I}}}_{{{g}_{7}}}}{{\theta }_{{{g}_{7}}}}\left( {{x}_{1}} \right) \right) \right)}_{g_{1}g_{8}} }  {{g}^{{{a}_{3}}{{a}_{2}}}}  \\ 
& \times {{{\hat{D}}}_{{{a}_{3}}}}\left( {{{{A}'}}_{{{a}_{3}},{{g}_{3}}}}\left( {{x}_{1}} \right) \right){{F}_{{{a}_{1}}{{a}_{2}},{{g}_{8}}}}\left( {{{{A}'}}_{{{a}_{k}},{{g}_{k}}}}\left( {{x}_{1}} \right) \right)=0. \\ 
\end{split}
\label{eq:kalibrovocnaja_invariantnost}
\end{equation}
where we denote this configurations with primes. In Eq.(\ref{eq:kalibrovocnaja_invariantnost}) the
parameters of a local gauge transformation denoted by $\theta_{g_{k}}(x_{1})$ and $\theta_{g_{k}}(x_{2})$ respectively,
$\hat{I}_{g_{k}}$ are generators of the adjoint representation of the gauge group. As seen from this expression, the inhomogeneous term that includes derivatives of the transformation parameter entering the equality as a product on the expression which is equal to zero as a result of the dynamical equations for the single-particle field $A_{a_{1},g_{1}}(x_{1})$. Therefore, only the first term remains in Eq.(\ref{eq:kalibrovocnaja_invariantnost}). Convoluting this term by indices $g_{1}$ and $g_{4}$ with matrices which are inverse of the matrices of the adjoint representation, we get the same system of equations for the field functions with primes as was written in Eq.(\ref{eq:Gluon_Lagrang_Eiler_naAx2}). Hereby one may say that in local gauge transformations, the left hand side of the system, Eq.(\ref{eq:Gluon_Lagrang_Eiler_naAx2}), transforms with respect to the tensor product of the two adjoint representations of the gauge group. Hence this left hand side of Eq.(\ref{eq:Gluon_Lagrang_Eiler_naAx2}) is a tensor of the Lorentz group with respect to indices $a_{1}$,$a_{4}$, and is also a tensor of local gauge transformations with respect to indices $g_{1}$,$g_{4}$. Furthermore we denote this tensor as $L_{a_{1}a_{4};g_{1}g_{4}}(x_{1},x_{2})$. For each these pairs of indices one may identically represent this tensor as a sum of the unit tensor, the antisymmetric tensor and the symmetric tensor with zero trace:
\begin{equation}
\begin{split}
{{L}_{{{a}_{1}}{{a}_{4}};{{g}_{1}}{{g}_{4}}}}\left( {{x}_{1}},{{x}_{2}} \right)=l\left( {{x}_{1}},{{x}_{2}} \right){{g}_{{{a}_{1}}{{a}_{4}}}}{{\delta }_{{{g}_{1}}{{g}_{4}}}}+\cdots .
\end{split}
\label{eq:rozclad_na_invariantni_pidprostori}
\end{equation}
Here we just pick out the term that is composed of unit tensors, the rest of the terms which are irrelevant for us denoted as ellipsis. Take into consideration that all such terms are linearly independent tensors Eq.(\ref{eq:Gluon_Lagrang_Eiler_naAx2}), we obtain:
\begin{equation}
\begin{split}
l\left( {{x}_{1}},{{x}_{2}} \right){{g}_{{{a}_{1}}{{a}_{4}}}}{{\delta }_{{{g}_{1}}{{g}_{4}}}}=0,
\end{split}
\label{eq:scalar_cast_L}
\end{equation}
or
\begin{equation}
\begin{split}
& {{g}^{{{a}_{1}}{{a}_{4}}}}{{\delta }_{{{g}_{1}}{{g}_{4}}}}{{A}_{{{a}_{4}},{{g}_{4}}}}\left( {{x}_{2}} \right){{g}^{{{a}_{3}}{{a}_{2}}}} \\ 
& \times {{{\hat{D}}}_{{{a}_{3}}}}\left( {{A}_{{{a}_{3}},{{g}_{3}}}}\left( {{x}_{1}} \right) \right){{F}_{{{a}_{1}}{{a}_{2}},{{g}_{1}}}}\left( {{A}_{{{a}_{k}},{{g}_{k}}}}\left( {{x}_{1}} \right) \right)=0. \\ 
\end{split}
\label{eq:abo}
\end{equation}
Given that $x_{1} \neq x_{2}$, the tensor $\delta_{g_{1}g_{2}}$ is not transformed into itself under the local transformation of Eq.(\ref{eq:kalibrovocnaja_invariantnost}) as was shown above. But because the local invariance of Eq.(\ref{eq:Gluon_Lagrang_Eiler_naAx2}) is achieved by a combination of the transformation of Eq.(\ref{eq:Gluon_Lagrang_Eiler_naAx2}) and a convolution with the components of matrices, that are inverse with respect to the matrices of the adjoint representation, then after such combination, the tensor $\delta_{g_{1}g_{2}}$ will already be transformed into itself. Moreover the Eqs.(\ref{eq:scalar_cast_L})-(\ref{eq:abo}) will preserve their form with respect to a local gauge transformation.

We symmetrize Eq.(\ref{eq:abo}) with respect to the variables $x_{1}$ and $x_{2}$. The obtained equation will include two tensors:
\begin{equation}
\begin{split}
& {{A}_{{{a}_{1}},{{g}_{1}}}}\left( {{x}_{1}} \right){{A}_{{{a}_{2}},{{g}_{2}}}}\left( {{x}_{2}} \right)={{\phi }_{{{a}_{1}}{{a}_{2}};{{g}_{1}}{{g}_{2}}}}\left( {{x}_{1}},{{x}_{2}} \right), \\ 
& {{A}_{{{a}_{1}},{{g}_{1}}}}\left( {{x}_{1}} \right){{A}_{{{a}_{2}},{{g}_{2}}}}\left( {{x}_{1}} \right) \\ 
& +{{A}_{{{a}_{1}},{{g}_{1}}}}\left( {{x}_{2}} \right){{A}_{{{a}_{2}},{{g}_{2}}}}\left( {{x}_{2}} \right)={{\chi }_{{{a}_{1}}{{a}_{2}};{{g}_{1}}{{g}_{2}}}}\left( {{x}_{1}},{{x}_{2}} \right). \\ 
\end{split}
\label{eq:dva_tenzori}
\end{equation}

For these tensors we have the relations:
\begin{equation}
\begin{split}
{{\chi }_{{{a}_{1}}{{a}_{2}};{{g}_{1}}{{g}_{2}}}}\left( {{x}_{1}},{{x}_{1}} \right)=2{{\phi }_{{{a}_{1}}{{a}_{2}};{{g}_{1}}{{g}_{2}}}}\left( {{x}_{1}},{{x}_{1}} \right).
\end{split}
\label{eq:spivvidnoshenna}
\end{equation}

Each of the tensors in Eq.(\ref{eq:dva_tenzori}) will be decomposed into invariant tensors with respect to Lorentz transformations
and to global internal transformations marking out the scalar part:
\begin{equation}
\begin{split}
& {{\phi }_{{{a}_{1}}{{a}_{2}};{{g}_{1}}{{g}_{2}}}}\left( {{x}_{1}},{{x}_{2}} \right)=\phi \left( {{x}_{1}},{{x}_{2}} \right){{g}_{{{a}_{1}}{{a}_{2}}}}{{\delta }_{{{g}_{1}}{{g}_{2}}}}+\cdots , \\ 
& {{\chi }_{{{a}_{1}}{{a}_{2}};{{g}_{1}}{{g}_{2}}}}\left( {{x}_{1}},{{x}_{2}} \right)=\chi \left( {{x}_{1}},{{x}_{2}} \right){{g}_{{{a}_{1}}{{a}_{2}}}}{{\delta }_{{{g}_{1}}{{g}_{2}}}}+\cdots . \\ 
\end{split}
\label{eq:Edi_tenzor}
\end{equation}
Note, that in this expression, we pick just these terms $\phi(x_{1},x_{2})$ and $\chi(x_{1},x_{2})$ because they are defined by Eq.(\ref{eq:Poznachenna_fi_xi}) and entering into the interaction Lagrangian Eq.(\ref{eq:L0_plus_Lint}) and (\ref{eq:LintProton}).

Now we will try to impose the solution that contains only terms selected in Eq.(\ref{eq:Edi_tenzor}) to Eq.(\ref{eq:abo}) which symmetrical with respect to $x_{1}$ and $x_{2}$. The rest of the terms in this solution are equal to zero. Note, that if, for the construction of the tensors of Eq.(\ref{eq:dva_tenzori}) one uses solutions of the single-particle equations, then the antisymmetric and the symmetric tensors with zero trace part, which are denoted in Eq.(\ref{eq:Edi_tenzor}) as ellipsis, in general are not equal to zero. Making them vanish means that from this moment we start considering a new pure two-particle field. Furthermore, for the two-particle gauge field we adopt the same sequence of the local transformations as in the previous sections. Namely, firstly we transform the whole tensors from Eq.(\ref{eq:dva_tenzori}), then we select from them a part that is proportional to the unit tensor. And by putting all other tensor parts equal to zero, except those that are selected in Eq.(\ref{eq:Edi_tenzor}), we obtain:

\begin{equation}
\begin{split}
 g^{a_{1}a_{2}}\frac{{{\partial }^{2}}\phi \left( {{x}_{1}},{{x}_{2}} \right)}{\partial x_{1}^{{{a}_{1}}}\partial x_{1}^{{{a}_{2}}}} & +  g^{a_{1}a_{2}}\frac{{{\partial }^{2}}\phi \left( {{x}_{1}},{{x}_{2}} \right)}{\partial x_{2}^{{{a}_{1}}}\partial x_{2}^{{{a}_{2}}}} \\ 
& -\frac{1}{2}{{g}^{2}}\phi \left( {{x}_{1}},{{x}_{2}} \right)\chi \left( {{x}_{1}},{{x}_{2}} \right)=0. \\ 
\end{split}
\label{eq:QCD_fi_xi}
\end{equation}
Furthermore, note that from Eq.(\ref{eq:spivvidnoshenna}) it follows that two-particle fields $\phi(x_{1},x_{2})$ and $\chi(x_{1},x_{2})$ should satisfy the requirement:
\begin{align}
\chi \left( {{x}_{1}},{{x}_{1}} \right)=2\phi \left( {{x}_{1}},{{x}_{1}} \right).
\label{eq:umova}
\end{align}
This requirement will be taken into account in the consideration of the solutions that correspond to two-particle fields.

Now we introduce new two-particle fields $a(x_{1},x_{2})$ and $b(x_{1},x_{2})$  instead of $\phi(x_{1},x_{2})$ and $\chi(x_{1},x_{2})$  using the following relation:
\begin{equation}
\begin{split}
& \phi \left( {{x}_{1}},{{x}_{2}} \right)=a\left( {{x}_{1}},{{x}_{2}} \right)-b\left( {{x}_{1}},{{x}_{2}} \right), \\ 
& \chi \left( {{x}_{1}},{{x}_{2}} \right)=a\left( {{x}_{1}},{{x}_{2}} \right)+b\left( {{x}_{1}},{{x}_{2}} \right). \\ 
\end{split}
\label{eq:Oznachenna_a_i_b}
\end{equation}

Within these variables, instead of the requirement Eq.(\ref{eq:umova}), we get:
\begin{align}
a\left( {{x}_{1}},{{x}_{1}} \right)=3b\left( {{x}_{1}},{{x}_{1}} \right).
\label{eq:umova1}
\end{align}

Taking into account Eq.(\ref{eq:Oznachenna_a_i_b}) instead of Eq.(\ref{eq:QCD_fi_xi}), we get:
\begin{equation}
\begin{split}
& {{g}^{{{a}_{1}}{{a}_{2}}}}\frac{{{\partial }^{2}}a\left( {{x}_{1}},{{x}_{2}} \right)}{\partial x_{1}^{{{a}_{1}}}\partial x_{1}^{{{a}_{2}}}}+{{g}^{{{a}_{1}}{{a}_{2}}}}\frac{{{\partial }^{2}}a\left( {{x}_{1}},{{x}_{2}} \right)}{\partial x_{2}^{{{a}_{1}}}\partial x_{2}^{{{a}_{2}}}}\\ 
& -\frac{1}{2}{{g}^{2}}{{a}^{2}}\left( {{x}_{1}},{{x}_{2}} \right)-\left( {{g}^{{{a}_{1}}{{a}_{2}}}}\frac{{{\partial }^{2}}b\left( {{x}_{1}},{{x}_{2}} \right)}{\partial x_{1}^{{{a}_{1}}}\partial x_{1}^{{{a}_{2}}}} \right.\\ 
& \left. +{{g}^{{{a}_{1}}{{a}_{2}}}}\frac{{{\partial }^{2}}b\left( {{x}_{1}},{{x}_{2}} \right)}{\partial x_{2}^{{{a}_{1}}}\partial x_{2}^{{{a}_{2}}}}-\frac{1}{2}{{g}^{2}}{{b}^{2}}\left( {{x}_{1}},{{x}_{2}} \right) \right)=0. \\ 
\end{split}
\label{eq:QCD_a_b}
\end{equation}

If we denote the left hand side of Eq.(\ref{eq:QCD_a_b}) that contains field $a(x_{1},x_{2})$ and its derivatives as $k(x_{1},x_{2})$ then:
\begin{equation}
\begin{split}
{{g}^{{{a}_{1}}{{a}_{2}}}}\frac{{{\partial }^{2}}a\left( {{x}_{1}},{{x}_{2}} \right)}{\partial x_{1}^{{{a}_{1}}}\partial x_{1}^{{{a}_{2}}}} & +{{g}^{{{a}_{1}}{{a}_{2}}}}\frac{{{\partial }^{2}}a\left( {{x}_{1}},{{x}_{2}} \right)}{\partial x_{2}^{{{a}_{1}}}\partial x_{2}^{{{a}_{2}}}} \\ 
& -\frac{1}{2}{{g}^{2}}{{a}^{2}}\left( {{x}_{1}},{{x}_{2}} \right)=k\left( {{x}_{1}},{{x}_{2}} \right), \\ 
\end{split}
\label{eq:k_ot_x1_x2}
\end{equation}
as seen from Eq.(\ref{eq:QCD_a_b}) the part that includes the $b(x_{1},x_{2})$ field should be equal to the same function
$k(x_{1},x_{2})$. We get the most simple problem by considering the case where $k(x_{1},x_{2})$ is equal to some constant. Let's consider the physical consequences which are driven by this particular case.

Therefore, we impose a partial solution for Eq.(\ref{eq:QCD_a_b}) that is defined by the following relations:
\begin{equation}
\begin{split}
{{g}^{{{a}_{1}}{{a}_{2}}}}\frac{{{\partial }^{2}}a\left( {{x}_{1}},{{x}_{2}} \right)}{\partial x_{1}^{{{a}_{1}}}\partial x_{1}^{{{a}_{2}}}} & +{{g}^{{{a}_{1}}{{a}_{2}}}}\frac{{{\partial }^{2}}a\left( {{x}_{1}},{{x}_{2}} \right)}{\partial x_{2}^{{{a}_{1}}}\partial x_{2}^{{{a}_{2}}}} -\frac{1}{2}{{g}^{2}}{{a}^{2}}\left( {{x}_{1}},{{x}_{2}} \right)=k, \\ 
{{g}^{{{a}_{1}}{{a}_{2}}}}\frac{{{\partial }^{2}}b\left( {{x}_{1}},{{x}_{2}} \right)}{\partial x_{1}^{{{a}_{1}}}\partial x_{1}^{{{a}_{2}}}} & +{{g}^{{{a}_{1}}{{a}_{2}}}}\frac{{{\partial }^{2}}b\left( {{x}_{1}},{{x}_{2}} \right)}{\partial x_{2}^{{{a}_{1}}}\partial x_{2}^{{{a}_{2}}}}- \frac{1}{2}{{g}^{2}}{{b}^{2}}\left( {{x}_{1}},{{x}_{2}} \right)=k, \\ 
\end{split}
\label{eq:chastkovij_rozvazok}
\end{equation}
where $k$ is some arbitrary constant.

Taking into account that the interaction Lagrangian Eq.(\ref{eq:LintProton}) depends only on the $a(x_{1},x_{2})$, we will examine only the first equation of Eq.(\ref{eq:chastkovij_rozvazok}). Assuming that due to similarity of these fields, all obtained results, will also be relevant for the field $b(x_{1},x_{2})$. Turning to the Jacobi coordinates Eq.(\ref{eq:Jacobi}), we get
\begin{equation}
\begin{split}
\frac{1}{2}{{g}^{{{a}_{1}}{{a}_{2}}}}\frac{{{\partial }^{2}}a\left( X,{{y}_{1}} \right)}{\partial {{X}^{{{a}_{1}}}}\partial {{X}^{{{a}_{2}}}}} & +2{{g}^{{{a}_{1}}{{a}_{2}}}}\frac{{{\partial }^{2}}a\left( X,{{y}_{1}} \right)}{\partial y_{1}^{{{a}_{1}}}\partial y_{1}^{{{a}_{2}}}} -\frac{1}{2}{{g}^{2}}{{a}^{2}}\left( X,{{y}_{1}} \right)=k. \\ 
\end{split}
\label{eq:QCDJacobi}
\end{equation}
Before we start examining this equation let's make it dimensionless. Considering it being dimensionless, we come straight forward to the conclusion that the gauge field tensor should have dimensions of $(1/l^{2})$, where $l$ is the unit of length. Then the gauge field has dimensions of $(1/l)$. Moreover, based on the form of the extended derivative, one may conclude that the coupling constant $g$ is dimensionless. Based on this, in Eq.(\ref{eq:QCDJacobi}) further we assume that the two-particle field $a(X,y_{1})$ was made dimensionless by $(1/l^{2})$, the constant $k$ $\textendash$ by $(1/l^{4})$, and the coordinates as well as the internal coordinates were made dimensionless by $l$. There is no special notation for dimensionless quantities, but we will assume that the following 
dimensionless procedure has already been done in Eq.(\ref{eq:QCDJacobi}). Thus we will further consider that all quantities are already dimensionless.

The nonhomogeneous equation, Eq.(\ref{eq:QCDJacobi}), has different features in cases $k\geq 0$ or $k~\textless ~0$. In case of $k~\textless~0$, the equation allows for partial solutions $a_{0}=-\left( 2\vert k\vert/g^{2} \right)^{1/2} $, or $a_{0}=\left( 2\vert k\vert/g^{2} \right)^{1/2}$ which may be reduced to constants. Then by introducing a new field  $a_{1}(X,y_{1})$ with the help of the relation $a(X,y_{1})=a_{0}+a_{1}(X,y_{1})$ one may obtain a homogeneous equation for this field.

In case of $k\geq 0$ in order to come to the homogeneous equation, we need at least one partial solution of Eq.(\ref{eq:QCDJacobi}). This solution should be a function of internal variables $a_{0}(y_{1})$. On the subset of Eq.(\ref{eq:Odnochasnist}) the equation for this function has the form:
\begin{equation}
\begin{split}
-\Delta_{\mathbf{y}_{1}}{{a}_{0}}\left( \mathbf{y}_{1} \right)-\frac{1}{4}{{g}^{2}}a_{0}^{2}\left(  \mathbf{y}_{1} \right)=k.
\end{split}
\label{eq:Rivnanna_dla_potencialu}
\end{equation}
Before we start analyzing features of this expression, let's examine this expression from a physical point of view. Representing the field $a(X,y_{1})$ as:
\begin{equation}
\begin{split}
a\left( X,  \mathbf{y}_{1} \right)={{a}_{0}}\left(  \mathbf{y}_{1} \right)+{{a}_{1}}\left( X,  \mathbf{y}_{1} \right),
\end{split}
\label{eq:a0_plus_a1}
\end{equation}
the same substitution we should make in the Lagrangian in Eq.(\ref{eq:LintProton}). But then in the full baryon Lagrangian $L=L_{0}+L_{\rm{int}}$, where $L_{0}$ is defined by Eq.(\ref{eq:Dirac3Lagrangian}) and $L_{\rm{int}}$ by Eq.(\ref{eq:LintProton}), the terms that include $a_{0}(\mathbf{y}_{1})$ may be grouped with terms that contain $3m$. In this way, one may include terms with $a_{0}(\mathbf{y}_{1})$ into $L_{0}$, at the same time terms with  $a(X,\mathbf{y}_{1})$ can be considered as the interaction Lagrangian. For this new $L_{0}$ Lagrangian one may find an approximated Euler-Lagrange equation and instead of Eq.(\ref{eq:Dirac3particls}) obtain:
\begin{equation}
\begin{split}
& i\gamma _{{{s}_{1}}{{s}_{2}}}^{a}\frac{\partial {{\Psi }_{{{s}_{2}}}}\left( X,\mathbf{y}_{1},\mathbf{y_{2}} \right)}{\partial {{X}^{a}}}\\ 
& -\left( 3m-\frac{3}{4m}{{\Delta }_{\mathbf{y}_{1}}} \right.-\frac{1}{m}{{\Delta }_{\mathbf{y_{2}}}} \\ 
& -\frac{{{g}^{2}}}{9m}{{a}_{0}}\left( \mathbf{y_{2}} \right)-\frac{{{g}^{2}}}{9m}{{a}_{0}}\left( \mathbf{y}_{1} - \frac{1}{2}\mathbf{y_{2}} \right) \\ 
& \left. -\frac{{{g}^{2}}}{9m}{{a}_{0}}\left( \mathbf{y}_{1}+\frac{1}{2}\mathbf{y_{2}} \right) \right){{\Psi }_{{{s}_{1}}}}\left( X, \mathbf{y}_{1},\mathbf{y_{2}} \right)=0. \\ 
\end{split}
\label{eq:Dirac_s_potencialom}
\end{equation}

Thus the function $\left(-a_{0}(\mathbf{y}_{1})\right)$ included into the internal Lagrangian as potential energy of quark interactions inside the hadron in non-relativistic approximation with accuracy up to terms with $g^{2}/9m$. Based on this we want to analyze features of the function $\left(a_{0}(\mathbf{y}_{1})\right)$ as a solution of Eq.(\ref{eq:Rivnanna_dla_potencialu}).

We turn Eq.(\ref{eq:Rivnanna_dla_potencialu}) into spherical variables and examine the simplest spherically symmetric solution of this equation. Denote $\vert\mathbf{y}_{1}\vert \equiv r$ and introduce a new unknown function $\left(a_{2}(\mathbf{y}_{1})\right)$ instead of $\left(a_{0}(\mathbf{y}_{1})\right)$, using the following relation:
\begin{equation}
\begin{split}
{{a}_{0}}\left( r \right)=\frac{{{a}_{2}}\left( r \right)}{r}.
\label{eq:a_2}
\end{split}
\end{equation}
Then for the function $\left(-a_{2}(r)\right)$ we have:
\begin{equation}
\begin{split}
\frac{{{d}^{2}}\left( -{{a}_{2}}\left( r \right) \right)}{d{{r}^{2}}}=kr+\frac{1}{4}{{g}^{2}}\frac{{{\left( -{{a}_{2}}\left( r \right) \right)}^{2}}}{r}.
\end{split}
\label{eq:Dif_ur_a_2}
\end{equation}

As was noted from Eq.(\ref{eq:a_2}) we get the function $\left(-a_{0}(r)\right)$ with finite value at $r=0$, also with finite derivative at this point if we apply the boundary conditions of Eq.(\ref{eq:Dif_ur_a_2}) as:
\begin{equation}
\begin{split}
{{\left. {{a}_{2}}\left( r \right) \right|}_{r=0}}=0,{{\left. \frac{d{{a}_{2}}\left( r \right)}{dr} \right|}_{r=0}}=C.
\end{split}
\label{eq:Granichni_umovi}
\end{equation}

The value of $C$ is not defined from the analyticity of the function $\left(-a_{0}(r)\right)$ and therefore it can be arbitrary. Thus let's analyze the features of the solution of Eq.(\ref{eq:Dif_ur_a_2}) with respect to the different values of $C$. The features of the function $\left(-a_{2}(r)\right)$ can be analyzed with the help of an expansion into a Taylor series taking into account that we are looking for the function $\left(-a_{2}(r)\right)$ at the manifold of functions that are analytic at the vicinity of the point $r=0$. Due to the boundary conditions Eq.(\ref{eq:Granichni_umovi}), the expansion of the function $\left(-a_{2}(r)\right)$ into a Taylor series can be started from the second order term. But if we substitute the expression:
\begin{equation}
\begin{split}
\left( -{{a}_{2}}\left( r \right) \right)=\sum\limits_{l=2}^{\infty }{{{q}_{l}}{{r}^{l}}},
\end{split}
\label{eq:Rozclad_Tejlor}
\end{equation}
into Eq.(\ref{eq:Dif_ur_a_2}), we get:
\begin{equation}
\begin{split}
& {{q}_{2}}=0,{{q}_{4}}=0,{{q}_{5}}=0,{{q}_{6}}=0, \\ 
& {{q}_{3}}=\frac{k}{6},{{q}_{7}}=\frac{{{g}^{2}}{{k}^{2}}}{2688},\ldots  \\ 
\end{split}
\label{eq:koeficienti}
\end{equation}
It is noticeable from these relations that Eq.(\ref{eq:Dif_ur_a_2}) has a nontrivial solution that satisfies the boundary conditions Eq.(\ref{eq:Granichni_umovi}). From these relations one may also note that in the case $k=0,C=0$ this equation will just have a trivial solution with the same boundary conditions. The case $k=0$ we will be discussed in detail later in the paper.

By integrating Eq.(\ref{eq:Dif_ur_a_2}) twice with the conditions of Eq.(\ref{eq:Granichni_umovi}) we get:
\begin{equation}
\begin{split}
& \left( -{{a}_{2}}\left( r \right) \right)=\frac{k}{6}{{r}^{3}} + \frac{1}{4}{{g}^{2}}\int\limits_{0}^{r}{d{{r}_{1}}}\int\limits_{0}^{{{r}_{1}}}{d{{r}_{2}}}\frac{\left( - a_{2}(r_{2})  \right)^{2}} {r_{2}}, \\ 
\end{split}
\label{eq:Kofainment1}
\end{equation}
whence, using the inalienability of the integrand in expression Eq.(\ref{eq:Kofainment1}) and relation Eq.(\ref{eq:a_2}) we obtain:
\begin{equation}
\begin{split}
\left( -{{a}_{0}}\left( r \right) \right)\ge \frac{k}{6}{{r}^{2}}.
\end{split}
\label{eq:Otcenca_potencialu}
\end{equation}
Thus, obtained from two-particle equations, the QCD potential of quark interactions, $\left(-a_{0}(r)\right)$, provides not only the possibility of existence of hadrons as bound particles, but also quark confinement. Note, that the case, $k~\textgreater~0, C=0$, which has just been discussed in addition to quark confinement also describes the features of  asymptotic freedom.

Now we examine the case $k~\textgreater~0, C~\textgreater~0$. By integrating Eq.(\ref{eq:Dif_ur_a_2}) twice we get:
\begin{equation}
\begin{split}
& \left( -{{a}_{2}}\left( r \right) \right)=Cr+\frac{k}{6}{{r}^{3}} + \frac{1}{4}{{g}^{2}}\int\limits_{0}^{r}{d{{r}_{1}}}\int\limits_{0}^{{{r}_{1}}}{d{{r}_{2}}}\frac{\left( - a_{2}(r_{2})  \right)^{2}} {r_{2}}. \\  
\end{split}
\label{eq:C_ne_nol}
\end{equation}
From this relation one can make an estimation of:
\begin{equation}
\begin{split}
\left( -{{a}_{2}}\left( r \right) \right)>Cr\ge 0,
\end{split}
\label{eq:ocenca}
\end{equation}
and as a consequence we get the inequality
\begin{equation}
\begin{split}
\left( -{{a}_{2}}\left( r \right) \right)>Cr+\frac{1}{6}\left( k+\frac{{{g}^{2}}{{C}^{2}}}{4} \right){{r}^{3}}.
\end{split}
\label{eq:nerivist}
\end{equation}
Thus, as seen from this relation, in case of $C~\textgreater~0$ the potential $\left(-a_{0}(r)\right)$ will ensure the existence of a bound state of quarks and confinement not only in case of $k~\textgreater~0$, but also at $k=0$. By considering the Taylor series at the vicinity of $r=0$, by analogy with Eq.(\ref{eq:Rozclad_Tejlor}) and Eq.(\ref{eq:koeficienti}) we get:
\begin{equation}
\begin{split}
& \left( -{{a}_{2}}\left( r \right) \right)\xrightarrow{r\to 0}Cr + \frac{1}{6}\left( k+\frac{g^{2}C^{2}}{4} \right){{r}^{3}}.\\ 
\end{split}
\label{eq:mal_r}
\end{equation}
Therefore, the conclusion about the description of the asymptotic freedom features remains valid in the case of $k~\textgreater~0, C\neq0$, as well as in the case of $k=0, C\neq0$.

We now analyze the existence of the bound states and confinement in the case $k~\textgreater~0, C~\textless~0$. We are interested in the asymptotic behavior of the function $\left(-a_{0}(r)\right)$, hence the behavior of $\left(-a_{2}(r)\right)$ at high $r$. In order to make this analysis, we rewrite Eq.(\ref{eq:ocenca}) as:
\begin{equation}
\begin{split}
& \left( -{{a}_{2}}\left( r \right) \right)=r\left( C+\frac{k}{6}{{r}^{2}} \right) +\frac{1}{4}{{g}^{2}}\int\limits_{0}^{r}{d{{r}_{1}}}\int\limits_{0}^{{{r}_{1}}}{d{{r}_{2}}}\frac{\left( - a_{2}(r_{2})  \right)^{2}} {r_{2}}. \\ 
\end{split}
\label{eq:C_lt_nol}
\end{equation}
The expression $\left( C+\frac{k}{6}{{r}^{2}} \right)$ at relatively high values of $r$ becomes positive. If we denote by $r_{0}$ the value of $r$ at which this expression is equal to some positive value $C_{0}$, then for $r~\textgreater~r_{0}$ we get:
\begin{equation}
\begin{split}
& \left( -{{a}_{2}}\left( r \right) \right)>{{C}_{0}}r +\frac{1}{4}{{g}^{2}}\int\limits_{{{r}_{0}}}^{r}{d{{r}_{1}}}\int\limits_{{{r}_{0}}}^{{{r}_{1}}}{d{{r}_{2}}} \frac{\left( - a_{2}(r_{2})  \right)^{2}} {r_{2}}>{C_{0}}r>0. \\ 
\end{split}
\label{eq:trojnoe_neravenstvo}
\end{equation}
As a result of this inequality we have the estimate:
\begin{equation}
\begin{split}
\left( -{{a}_{2}}\left( r \right) \right)>{{C}_{0}}r+\frac{1}{24}{{g}^{2}}{{\left( {{C}_{0}} \right)}^{2}}{{r}^{3}}.
\end{split}
\label{eq:konfaiment_C_lt_0}
\end{equation}
Thus, as seen, the conclusion about the description of the quark confinement and asymptotic freedom for the case $k~\textgreater~0$ took place on arbitrary value of the constant $C$.
\begin{figure*}[t]
	\centering
	\subfigure[]{
		\includegraphics[scale=0.245]{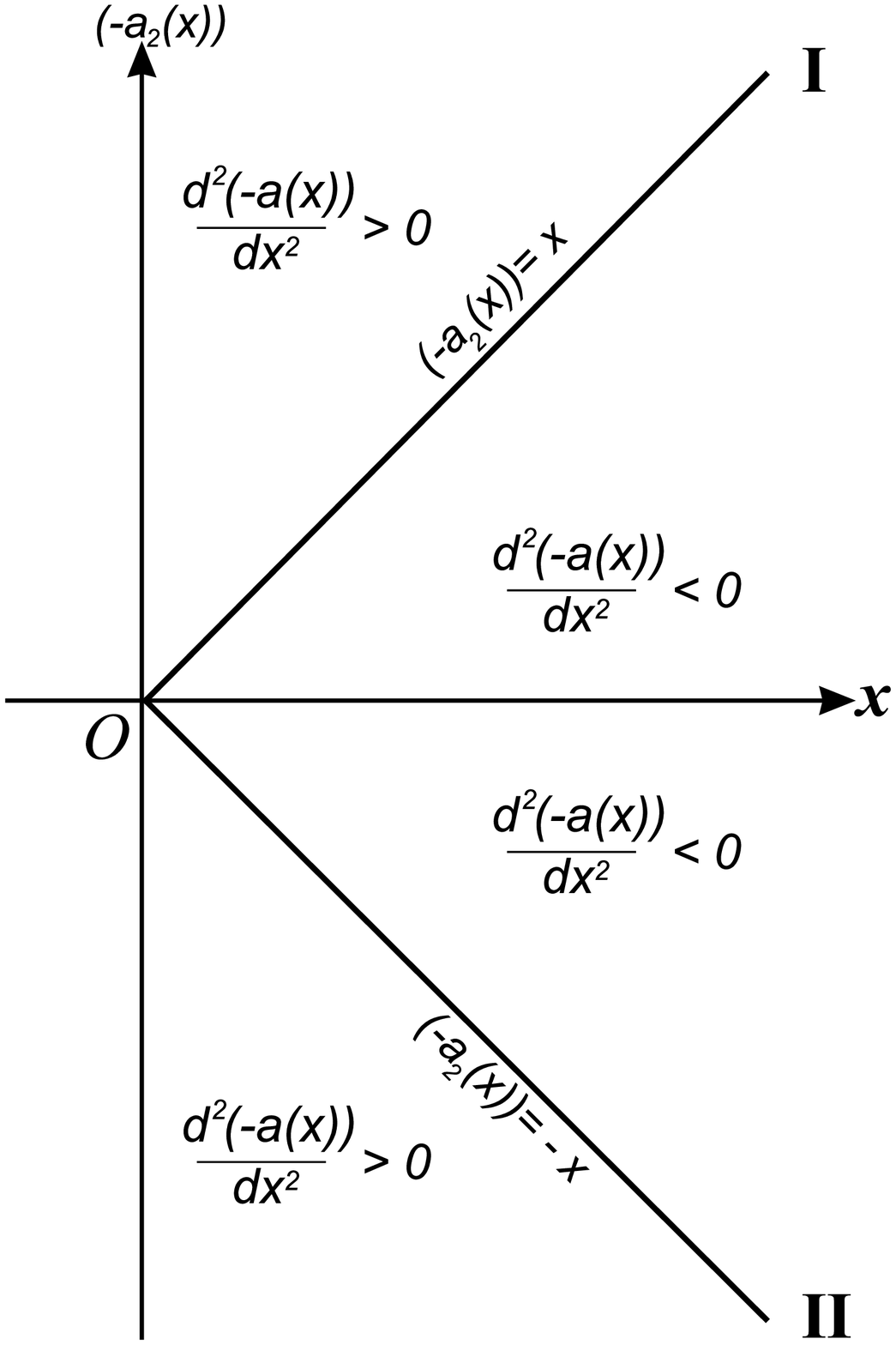}
	}
	\subfigure[]{
		\includegraphics[scale=0.245]{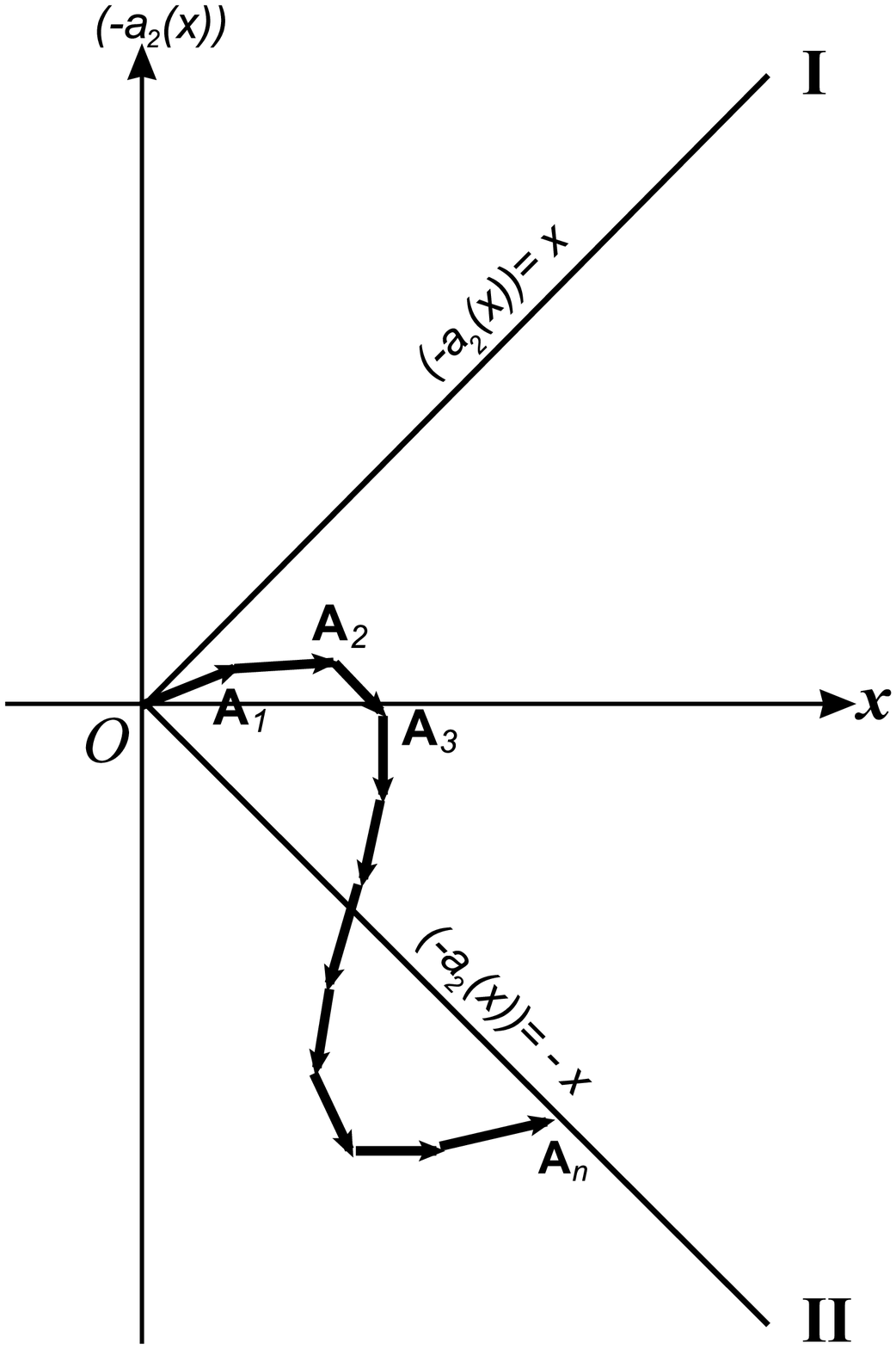}
	}
	\subfigure[]{
		\includegraphics[scale=0.245]{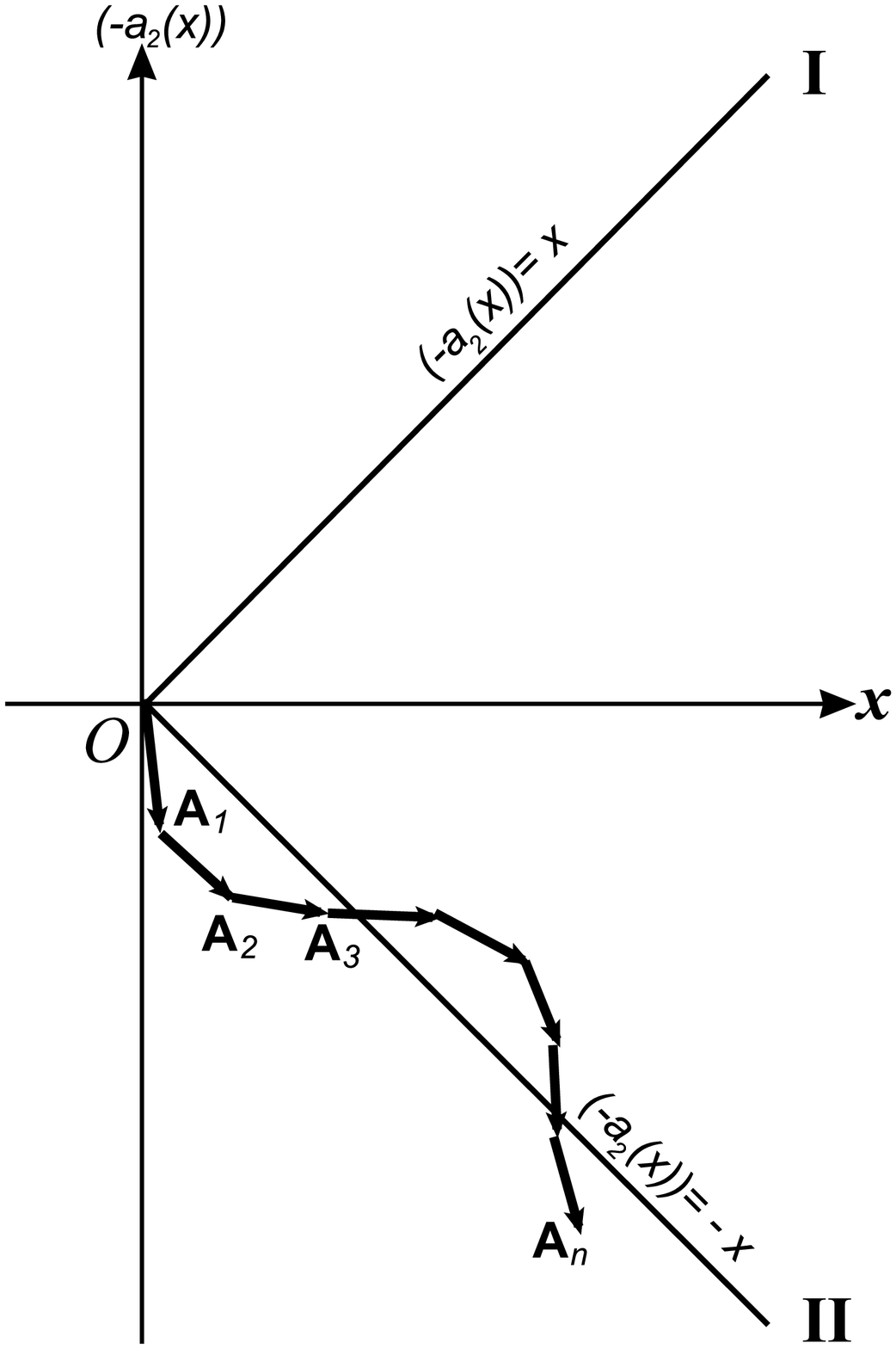}
	}
	\subfigure[]{
		\includegraphics[scale=0.245]{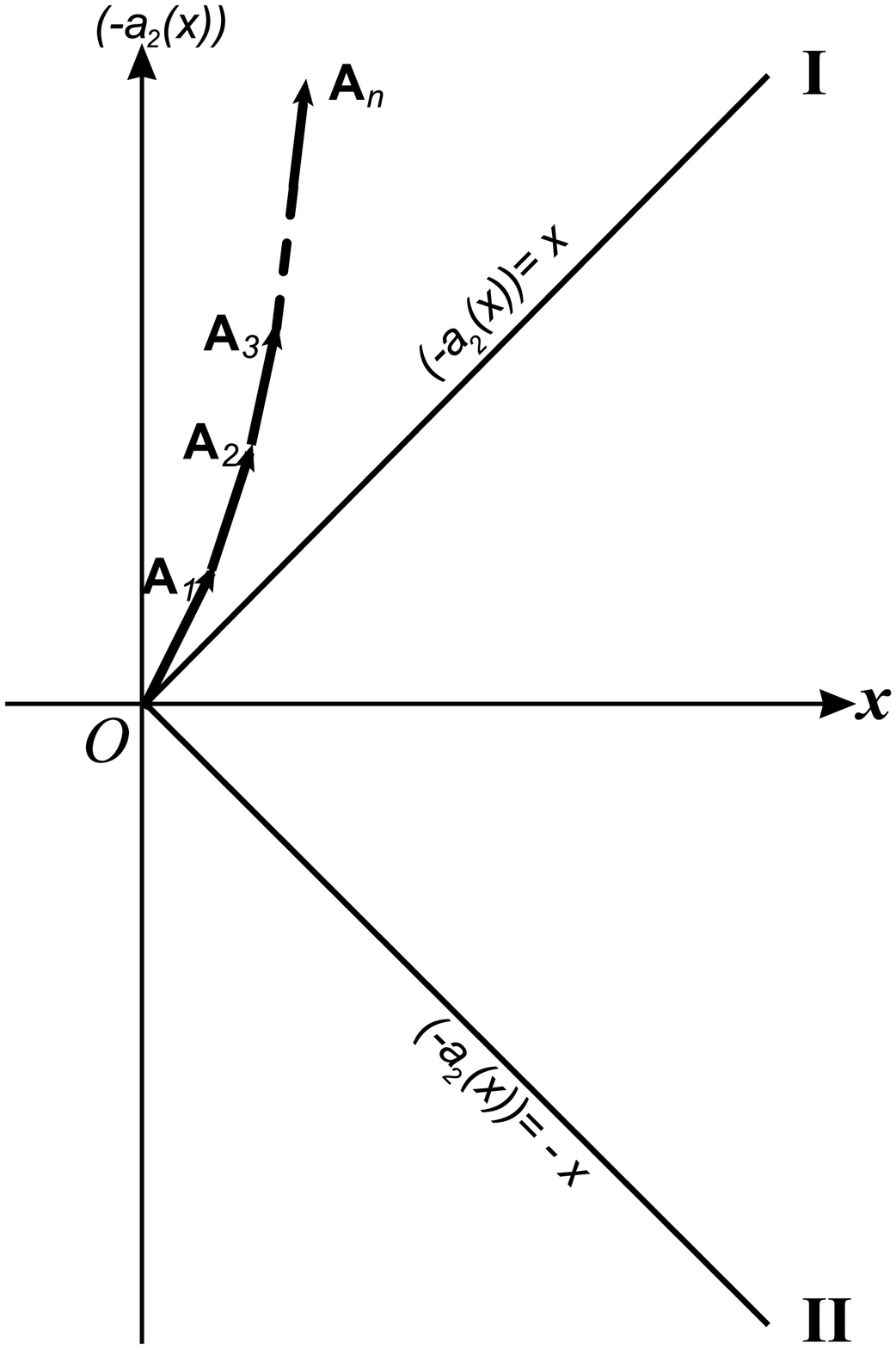}
	}
	\caption{
		(a) Dividing the phase half-plane into sectors with constant sign of the second derivative $\frac{d^{2}(-a_{2}(x))}{dx^{2}}$
		(b) Typical phase trajectory in the case of $-1\textless C^{\prime}\textless 1$. 
		(c) Analysis of the behavior of the solution of Eq.(\ref{eq:novij_vid}) which satisfies the conditions
		$(-a_{2}(x))\vert_{x=0}=0, \frac{d(-a_{2}(x))}{dx}\vert_{x=0}=C^{\prime}\textless -1$
		on the phase plane.
		(d) Analysis of the behavior of the solution of Eq.(\ref{eq:novij_vid}) which satisfies the conditions $C^{\prime}\textgreater -1$ on the phase plane.
	}
	\label{fig:PhasePortrait}
\end{figure*}

For the case $k~\textless~0$, features of the solutions are much more diverse, therefore we analyze these features on the phase plane. For this we introduce a new variable $x$ into Eq.(\ref{eq:Dif_ur_a_2}) with the help of the relation:
\begin{equation}
\begin{split}
r=\frac{g}{2\sqrt{\left| k \right|}}x. 
\end{split}
\label{eq:nova_zminna}
\end{equation}
Then we rewrite this expression in the form:
\begin{equation}
\begin{split}
& \frac{{{d}^{2}}\left( -{{a}_{2}}\left( x \right) \right)}{d{{x}^{2}}} = \left( \frac{{{g}^{3}}}{8\sqrt{\left| k \right|}} \right)\frac{\left( \left( -{{a}_{2}}\left( x \right) \right)-x \right)\left( \left( -{{a}_{2}}\left( x \right) \right)+x \right)}{x}. \\ 
\end{split}
\label{eq:novij_vid}
\end{equation}
It is seen from Eq.(\ref{eq:novij_vid}) that the equation has two trivial solutions $\left(-a_{2}(x)\right)=-x$ and $\left(-a_{2}(x)\right)=x$, which differ by boundary conditions:
\begin{equation}
\begin{split}
{{\left. \left( -{{a}_{2}}\left( x \right) \right) \right|}_{x=0}}=0,~~{{\left. \frac{d\left( -{{a}_{2}}\left( x \right) \right)}{dx} \right|}_{x=0}}\equiv {C}'=\pm 1.
\end{split}
\label{eq:Gr_umo_plu_min_odin}
\end{equation}
Here we use the notation $C^{\prime}$ for the first order derivative taken with respect to $\left(-a_{2}(x)\right)$ when $x=0$. Furthermore, we examine the behavior of the rest of the solutions with respect to the value of $C^{\prime}$. Half-lines $\left(-a_{2}(x)\right)=-x, x\geq 0$ and $\left(-a_{2}(x)\right)=x, x\geq 0$ split the phase half-space $x\geq 0$ into regions where the sign of the second order derivative $d^{2}\left(-a_{2}(x)\right)/dx^{2}$ is defined as shown in Fig.\ref{fig:PhasePortrait}(a).

For further analysis, it is better to use a mechanical analogy. We will call, the second derivative an ``acceleration", the first one $\textendash$ ``velocity", the value $C^{\prime}$ we will call the ``initial velocity". The function $\left(-a_{2}(x)\right)$ will be called the coordinate and its arguments we call as ``time". Therefore, we have the following analogy, there are two objects $\mathbf{I}$ and $\mathbf{II}$ which satisfy the solutions shown in Fig.\ref{fig:PhasePortrait} as straight lines. Further we analyze the behavior of the arbitrary solution, the object $\mathbf{III}$, with respect to these two. Thus at the beginning when time equals zero, there are three objects that start to move. The objects $\mathbf{I}$ and $\mathbf{II}$ are moving with constant velocity 1 and (-1) respectively. Movement of the object $\mathbf{III}$ is described by Eqs.(\ref{eq:novij_vid}) with an initial coordinate that is equal to zero and with initial velocity $C^{\prime}$. 

Examine the possible movement of the object $\mathbf{III}$ with respect to the initial velocity $C^{\prime}$. First, we analyze the case when $-1~\textless~C^{\prime}~\textless~1$, then object $\mathbf{III}$ gets into region of the negative acceleration (see, segment $OA_{1}$ on Fig.\ref{fig:PhasePortrait}(b)). In order to change sign of the acceleration of object $\mathbf{III}$, as seen from Eqs.(\ref{eq:novij_vid}), it should outrun object $\mathbf{I}$ or $\mathbf{II}$. Given that the initial velocity of $\mathbf{III}$ was lower with respect to the velocity of $\mathbf{I}$ and object $\mathbf{III}$, is located in the sector of the negative acceleration and it will never overtake object $\mathbf{I}$. This leads to the fact that the velocity of $\mathbf{III}$ will decrease (see, segments $A_{1}A_{2},A_{2}A_{3},...$ on Fig.\ref{fig:PhasePortrait}(b)). The acceleration of $\mathbf{III}$ will remain negative and nonzero until it overtakes object $\mathbf{II}$. This means that it will surely catch up the object $\mathbf{II}$, since in another case object $\mathbf{III}$ will stay in the sector of negative acceleration forever. This in turn means that nothing can prevent that the velocity of object $\mathbf{III}$ becomes negative and large with respect to the velocity of $\mathbf{II}$ by absolute value, which in turn leads to the fact that object $\mathbf{III}$ will overtake object $\mathbf{II}$. Once this happens, the object will fall into a region of  positive acceleration. The velocity of this object will start decreasing by absolute value until it will be overtaken by object $\mathbf{II}$. Furthermore, this situation will be repeated continuously, as shown in Fig.\ref{fig:PhasePortrait}(b).

In Fig.2(a) is shown the result of the numerical solution of Eq.\ref{eq:novij_vid} for the case $C^{\prime}=0.5, g=10, k=-5$. The corresponding potential is shown in Fig.3(solution (a)).

Results shown in Fig.2(a) and Fig.3(solution (a)) correspond to the value of $\approx 55.9$ for the coefficient $\left(\frac{g^{3}}{8\sqrt{\vert k \vert}} \right)$ in Eq.(\ref{eq:novij_vid}). If one will increase this coefficient by a factor of 1000, the solutions $(-a_{2}(x))$ and $(-a_{2}(x))=-x$ will almost match each other, as shown in Fig.2(d).

If one will decrease this coefficient, for instance $\left(\frac{g^{3}}{8\sqrt{\vert k \vert}} \right)\approx 5.59\times 10^{-5}$, we will get a much longer process as shown in Fig.2(b).

Now let's examine the features of the solution that corresponds to the initial condition $C^{\prime}~\textless ~-1$. In this case when the object $\mathbf{III}$ will outrun object $\mathbf{II}$, it will fall into the region of positive acceleration (see, segments $OA_{1}$ in Fig.\ref{fig:PhasePortrait}(c)). This in turn will lead to the fact that the velocity of the object $\mathbf{III}$ with respect to the object $\mathbf{II}$ starts to decrease, and the object $\mathbf{II}$ will begin to overtake the object $\mathbf{III}$ (see, segments $A_{1}A_{2},A_{2}A_{3},...$ on Fig.\ref{fig:PhasePortrait}(c)). This process will not stop until $\mathbf{II}$ will not outrun the object $\mathbf{III}$. This outrun will happen anyway. But afterwards, the object $\mathbf{III}$ will fall into the region of negative acceleration and its velocity with time will again become higher in absolute value with respect to the velocity of the object $\mathbf{II}$. This situation will lead to the fact that object $\mathbf{II}$ will again overtake object $\mathbf{III}$. This situation will be repeated continuously, as this shown in Fig.\ref{fig:PhasePortrait}(c). This discussion is confirmed by the results of the numerical solutions of Eq.(\ref{eq:novij_vid}). A typical example of the solution for boundary conditions that are examined is shown in Fig.2(c). The corresponding potential $\left(-a_{0}(r)\right)$ is shown in Fig.3(solution (b)).

In case of $C^{\prime}~\textgreater~1$, the object $\mathbf{III}$ will immediately outrun the object $\mathbf{I}$ and will fall into the region of the positive acceleration as shown in Fig.\ref{fig:PhasePortrait}(d). This will lead to an increase of the velocity of the object $\mathbf{III}$, which already was high with respect to the velocity of the object $\mathbf{I}$. In this case the object $\mathbf{III}$ will never fall into the region of negative acceleration. A typical result of a numerical solution for this case, $C^{\prime}~\textgreater~1$, is shown in Fig.\ref{fig:x_vs_a2x_Graph9_comb}(left). 

For small values of the coupling constant we get qualitatively the same behavior for the solution (see Fig.\ref{fig:x_vs_a2x_Graph9_comb}(right)) that differs from the solution shown in Fig.\ref{fig:x_vs_a2x_Graph9_comb}(left) just by numerical values.

Thus in the case of $C^{\prime}~\textgreater~1$, with respect to the two previous cases, we again have a quark confinement and asymptotic freedom. Fig.\ref{fig:x_vs_a2x_Graph9_comb}(left) shows that at large $x$, the function $\left(-a_{2}(x)\right)$ is growing much faster than $x$. Therefore, in the limit when $r$ approaches infinity, the potential $\left(-a_{0}(x)\right)$ or $\left(-a_{0}(r)\right)$ will also go to infinity. This means that in order to set apart two quarks to infinity, one may need an infinite potential energy, which leads to confinement. At same time one may note from Fig.\ref{fig:x_vs_a2x_Graph9_comb}(left) that in the limit of small $x$, the $\left(-a_{2}(x)\right)=x$ and hence $\left(-a_{0}(x)\right) = \left(-a_{2}(x)\right)/x$ will approach some constant. But if the potential energy approach to a constant, then the gradient, which is equal to force, will approach to zero. Hence at small distances, quarks do not interact, which corresponds to asymptotic freedom.

Considering the equations and the features of their solutions allows us to reply one question: What is the two-particle gauge field from a physical point of view? To reply it, we return to Eq.(\ref{eq:QCDJacobi}) and examine its solution on the subset of simultaneity, Eq.(\ref{eq:Odnochasnist}). This equation is nonhomogeneous. If we want to quantize the fields $a\left(X, \mathbf{y}_{1} \right)$ and $b \left(X, \mathbf{y}_{1} \right)$, we have to start from homogenues equations. This can be done by substituting Eq.\ref{eq:QCDJacobi}  into Eq.(\ref{eq:a0_plus_a1}), where $a_{0} \left( \mathbf{y}_{1} \right)$ is one of the solutions of Eq.(\ref{eq:Rivnanna_dla_potencialu}) and $a_{1} \left( X, \mathbf{y}_{1} \right)$ is a new unknown function. Therefore, we get the following equation for this unknown function:			
\begin{equation}
\begin{split}
& {{g}^{{{a}_{1}}{{a}_{2}}}}\frac{{{\partial }^{2}}{{a}_{1}}\left( X, \mathbf{y}_{1} \right)}{\partial {{X}^{{{a}_{1}}}}\partial {{X}^{{{a}_{2}}}}}+\left( -4{{\Delta }_{ \mathbf{y}_{1} }}{{a}_{1}}\left( X, \mathbf{y}_{1} \right) \right. \\ 
& \left. +2{{g}^{2}}\left( -{{a}_{0}}\left( \mathbf{y}_{1} \right) \right){{a}_{1}}\left( X, \mathbf{y}_{1} \right) \right) -{{g}^{2}}{{\left( {{a}_{1}}\left( X, \mathbf{y}_{1} \right) \right)}^{2}}=0. \\ 
\end{split}
\label{eq:fi3_gluon}
\end{equation}

To describe the field $a_{1} \left( X, \mathbf{y}_{1} \right)$, we will apply methods of perturbation theory. Thus we will drop the nonlinear terms from Eq.(\ref{eq:fi3_gluon}) and afterwards we get the equation that is generated by the action:
\begin{equation}
\begin{split}
S &=\int d^{4}X d\mathbf{y}_{1} \left( \frac{1}{2}{{g}^{{{a}_{1}}{{a}_{2}}}}\frac{\partial {{a}_{1}}\left( X,\mathbf{y}_{1} \right)}{\partial {{X}^{{{a}_{1}}}}}\frac{\partial {{a}_{1}}\left( X,\mathbf{y}_{1} \right)}{\partial {{X}^{{{a}_{2}}}}} \right. \\ 
& \left. -2\frac{\partial {{a}_{1}}\left( X,\mathbf{y}_{1} \right)}{\partial y_{1}^{b}}\frac{\partial {{a}_{1}}\left( X,\mathbf{y}_{1} \right)}{\partial y_{1}^{b}} - \frac{g^{2}}{2}(-a_{0}( \mathbf{y}_{1})) {{\left( {{a}_{1}}\left( X,\mathbf{y}_{1} \right) \right)}^{2}} \right). \\ 
\end{split}
\label{eq:Dija_a1}
\end{equation}
Noether's theorem for the field with such action leads to the following expression for the energy of the field $a_{1} \left( X, \mathbf{y}_{1} \right)$ at zeroth order in perturbation theory:
\begin{equation}
\begin{split}
{{P}_{0}}&=\frac{1}{2}\int{d{{X}^{1}}d{{X}^{2}}d{{X}^{3}}}d\mathbf{y}_{1} \\ 
& \times \left( \sum\limits_{b=0}^{3}{{{\left( \frac{\partial {{a}_{1}}\left( X, \mathbf{y}_{1} \right)}{\partial {{X}^{b}}} \right)}^{2}}} + 4\sum\limits_{b=1}^{3}{{{\left( \frac{\partial {{a}_{1}}\left( X,\mathbf{y}_{1} \right)}{\partial y_{1}^{b}} \right)}^{2}}} \right.\\ 
& \left. + {{g}^{2}}\left( -{{a}_{0}}\left( \mathbf{y}_{1} \right) \right){{\left( {{a}_{1}}\left( X,\mathbf{y}_{1} \right) \right)}^{2}} \right). \\ 
\end{split}
\label{eq:P0_a1}
\end{equation}
As can be seen from this expression, the requirement of finite energy leads to a quite fast approach to zero of all possible realizations of the field $a_{1} \left( X, \mathbf{y}_{1} \right)$ when $\vert \mathbf{y}_{1} \vert$ is approaching infinity in the case when $\left(-a_{0}(\mathbf{y}_{1})\right)$ preserves quark confinement. Thus, the same function $\left(-a_{0}(\mathbf{y}_{1})\right)$ that in Eq.(\ref{eq:Dirac_s_potencialom}) provides quark confinement in equation for the zeroth order approximation for the field $a_{1} \left( X, \mathbf{y}_{1} \right)$ is providing gluon confinement:
\begin{equation}
\begin{split}
& {{g}^{{{a}_{1}}{{a}_{2}}}}\frac{{{\partial }^{2}}{{a}_{1}}\left( X,\mathbf{y}_{1} \right)}{\partial {{X}^{{{a}_{1}}}}\partial {{X}^{{{a}_{2}}}}}\\ 
& +\left( -4\Delta_{\mathbf{y}_{1}} a_{1}\left( X,\mathbf{y}_{1} \right) + 2{{g}^{2}}\left( -{{a}_{0}}\left( \mathbf{y}_{1} \right) \right){{a}_{1}}\left( X,\mathbf{y}_{1} \right) \right)=0, \\ 
\end{split}
\label{eq:Nol_nab_glu}
\end{equation}

Eq.(\ref{eq:Nol_nab_glu}) is similar to Eq.(\ref{eq:LagEiler}) or (\ref{eq:KGFInternal}) with internal variables but with the minor difference that the operator
\begin{equation}
\begin{split}
& {{\left( {{{\hat{H}}}^{\text{internal}}} \right)}^{2}}\left( {{a}_{1}}\left( X, \mathbf{y}_{1} \right) \right) =\\ 
& -4 \Delta_{\mathbf{y}_{1}}{{a}_{1}}\left( X, \mathbf{y}_{1} \right)+{{g}^{2}}\left( -{{a}_{0}}\left( \mathbf{y}_{1} \right) \right){{a}_{1}}\left( X, \mathbf{y}_{1} \right), \\ 
\end{split}
\label{eq:H_vnut_kv}
\end{equation}
which is similar to the Hamiltonian of the two-particle system here is the square of the Hamiltonian of the internal system. Indeed if one will turn from the field $a_{1} \left( X, \mathbf{y}_{1} \right)$ to its Fourier representation using the coordinates $X^{a},~a=0,1,2,3$:
\begin{equation}
\begin{split}
{{a}_{1}}\left( X, \mathbf{y}_{1} \right) =\frac{1}{{{\left( 2\pi  \right)}^{{3}/{2}\;}}}\int{{{d}^{4}}X}{{a}_{1}}\left( p, \mathbf{y}_{1} \right) {\rm{e}}^{ i{{p}_{a}}{{X}^{a}} }, \\ 
\end{split}
\label{eq:Furrie}
\end{equation}
then we get the equation:
\begin{equation}
\begin{split}
{{\left( {{{\hat{H}}}^{\text{internal}}} \right)}^{2}}\left( {{a}_{1}}\left( p,\mathbf{y}_{1} \right) \right)=\left( {{g}^{{{a}_{1}}{{a}_{2}}}}{{p}_{{{a}_{1}}}}{{p}_{{{a}_{2}}}} \right){{a}_{1}}\left( p, \mathbf{y}_{1} \right). 
\end{split}
\label{eq:Sobst_zna}
\end{equation}
Thus, as seen from this equation, the eigenvalues of the operator, Eq.(\ref{eq:H_vnut_kv}), are equal to the square of the internal energy of the two-gluon particle for different internal states of this particle. Indeed, in the case of the potential $\left(-a_{0}(\mathbf{y}_{1})\right)$ that provides confinement, one may set boundary conditions for Eq.(\ref{eq:Dif_ur_a_2}) in a way, that all eigenvalues of the operator, Eq.(\ref{eq:H_vnut_kv}), will be positive. However, as follow from the previous discussion, this is true for cases where $k~\textgreater~0,C\geq0$ or $k=0, C~\textgreater~0$. Given Eq.(\ref{eq:Sobst_zna}), this means that the two-gluon particle has nonzero mass and there exists of a rest reference frame for it. As follow from Eq.(\ref{eq:Sobst_zna}), the eigenvalues of the operator in Eq.(\ref{eq:H_vnut_kv}) is equal to the square of the energy of the two-gluon particle in this reference frame.

From a formal point of view this situation looks similar to the one that was considered in Eq.(\ref{eq:LagEiler}), but with one significant difference. In Eq.(\ref{eq:LagEiler}) we have contribution from terms with $(2m)^{2}$, that comes from the mass of bound particles. Thus one may assume that the eigenvalues of a sum of the kinetic energy and potential energy operators, that are in a bound state, are small with respect to the total rest energy of these particles. In this way one may just keep linear terms with respect to these operators in the square of the internal Hamiltonian. But in the current case we are considering the bound state of two massless gluons, which can be noticed from Eq.(\ref{eq:H_vnut_kv}) due to absence of a term similar to $(2m^{2})$ in Eq.(\ref{eq:LagEiler}). Thus we can no longer assume that the eigenvalues of the operator in Eq.(\ref{eq:H_vnut_kv}) are small additive terms to something, so in order to examine the internal Hamiltonian of the two-gluon particle we have to take the square root of the operator in Eq.(\ref{eq:H_vnut_kv}).

A formal definition of this square root does not pose any problems, since in case of the confinement operator, Eq.(\ref{eq:H_vnut_kv}), it has just a discrete spectrum of eigenvalues. By choosing the system of eigenvalues that corresponds to this operator as a basis, on may represent it as a matrix. This matrix in this basis will be diagonal. On the main diagonal it will contain the eigenvalues of this operator. As was noticed above by choosing appropriate boundary conditions, Eq.(\ref{eq:Granichni_umovi}), they could be positive. If we will change these eigenvalues by the square root of them, then we get a matrix that would be a self-conjugate operator due to the realness of values of the square root. Therefore this operator can be used to define the operator $\hat{H}^{\rm{internal}}=\sqrt{\left( \hat{H}^{\rm{internal}} \right)^{2}}$. After the quantization procedure the sign in front this square root could be interpreted in a natural way by considering corresponding coefficients as creation and annihilation operators of the two-gluon particles.

Nevertheless, there is no need for an arbitrary definition of the operator $\hat{H}^{\rm{internal}}$, since as seen from the previous discussions all relations for the free field $a_{1}\left(X, \mathbf{y}_{1} \right)$, as well as for the field that interacts with others fields include especially the operator of Eq.(\ref{eq:H_vnut_kv}). Thus one may say that dynamics of the two-particle gauge field
is defined not by the internal Hamiltonian but by the square of it. Analysis of such kind of ``weird" situation allows one to make a conclusion about the nature of the two-particle gluon field. Since the operator $\left( -4\Delta_{\mathbf{y}_{1}}  \right)$ that is included in Eq.\ref{eq:H_vnut_kv} may be written as:
\begin{align}
-4\Delta_{\mathbf{y}_{1}}=-2\left( \Delta_{\mathbf{x}_{1}}+ \Delta_{\mathbf{x}_{2}} \right)+ \Delta_{\mathbf{X}}.
\label{eq:delta4x1x2}
\end{align}
We will consider the two-gluon particle in their center mass reference frame. Then the eigenvalue of $\Delta_{\mathbf{X}}$ is equal to zero, the rest of the operators could be considered as operators of the square of the energy each of the gluons:
\begin{equation}
\begin{split}
\hat{E}_{1}^{2}=-2\Delta_{\mathbf{x}_{1}},\\
\hat{E}_{2}^{2}=-2\Delta_{\mathbf{x}_{2}},\\
\end{split}
\label{eq:E_1_2}
\end{equation}
The operator $\left( -g^{2}a_{0}(\mathbf{y}_{1})  \right)$ describes the interaction between gluons. Due to this interaction the internal state of the two-gluon particle is not an eigenstate anymore, neither of energy nor of momentum of these gluons. But we can discuss the mean values of those quantities. Then from the definition of Eq.(\ref{eq:E_1_2}) we have:
\begin{align}
\left\langle E_{a}^{2} \right\rangle =2\left\langle  \mathbf{p}_{a}^{2} \right\rangle ,a=1,2.
\label{eq:seredni_znachenna}
\end{align}
If we consider the ``initial gluon", namely those that appear in QCD at the zero order of pertubation theory, then one will leave just linear terms in the equation. When QCD equations coincide with well-known QED equations, then due to masslessness of this gluon the relation between eigenvalues of the energy, $E$, and momentum, $\mathbf{p}$, in a state that is an eigenstate for both of these quantities, has the form $E=\vert\mathbf{p}\vert$. Then for the square of the average values in the noneigenstate for these quantities, we get exactly Eq.(\ref{eq:seredni_znachenna}). The factor of two appears in this relation due to two polarization states of the ``initial gluon". Therefore, masslessness of the ``initial gluon" leads to the fact that the square of the average momentum determines the square of the average energy of this particle, but not the average energy. This was shown previously for non-relativistic massive particles.
\begin{figure}[t]
	\centering
	\includegraphics[scale=0.35]{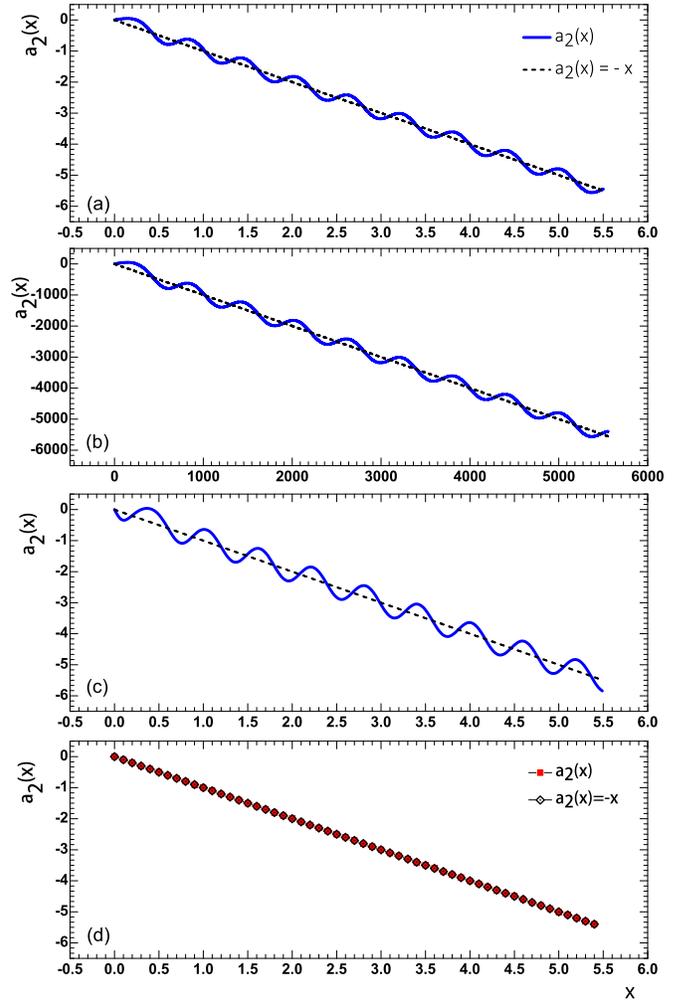}
	\caption{
		(a) Results of the numerical solution of Eq.(\ref{eq:novij_vid}) for $C^{\prime}=0.5, g=10, k=-5$;
		(b) for case of $\left( \frac{g^{3}}{8\sqrt{\vert k \vert}} \right)\approx 5.59\times 10^{-5}$;
		(c) for case of $C^{\prime}=-5.5, g=10, k=-5$.
		(d) same as pad (a) but the coefficient $\left( \frac{g^{3}}{8\sqrt{\vert k \vert}} \right)$ has been increase by a factor of 1000, so $\approx 55901.7$.		
	}
	\label{fig:ResNumericalSolutions}
\end{figure}

One may conclude from the above discussion, that two interacting gluons which form the two-gluon particle stay massless, even if their properties are significanly different from the properties of the ``initial gluon" from pertubation theory. Each of these gluons interact with one another and are in the state that is located in space, hence not in the eigenstate with respect to momentum. Due to this issue the average energy of the massless gluon in the examined state is not zero. Thus if we consider the possibility of creating a new gluon (or several gluons) in this reference frame, then it does not matter if it is massless because this creation will require nonzero energy (similar to the case when the gluon is massive). So, if the interaction of the two-gluon particle with another field cannot provide the required energy, then a new gluon will be not created, and the system will stay a two-gluon system. This makes sense for the consideration of the two-gluon field that describes the bound states of the two interactions between each massless gluons. Candidates for such two-gluon bound states are being searched for in the experiments \cite{Nerling:2011, Lamsa:2011} as well as in theory \cite{Frasca:2010}.

We can describe the internal state of the two-gluon particle by searching for eigenvalues of Eq.(\ref{eq:Sobst_zna}) for the operator in Eq.(\ref{eq:H_vnut_kv}). At this stage we can satisfy the condition, Eq.(\ref{eq:umova}), or Eq.(\ref{eq:umova1}). Using Jacobi coordinates, this requirement looks as:
\begin{align}
a\left( X,\mathbf{y}_{1}=0 \right)=3b\left( X,\mathbf{y}_{1}=0 \right).
\label{eq:umova_Jacobi}
\end{align}
As may see from Eq.(\ref{eq:chastkovij_rozvazok}), the fields $a \left(X, \mathbf{y}_{1} \right)$ and $b \left(X, \mathbf{y}_{1} \right)$ are satisfy the same equations. However, one can apply the different boundary conditions for them. Let's show that Eq.(\ref{eq:umova_Jacobi}) determines these boundary conditions.

Due to similarity of the equations for the field $b \left(X, \mathbf{y}_{1} \right)$ one can consider just a partial solution $b_{0} \left(\mathbf{y}_{1} \right)$ that is analog to $a_{0} \left(\mathbf{y}_{1} \right)$ and fulfill Eq.(\ref{eq:Rivnanna_dla_potencialu}) but using different boundary conditions. Namely if one will define $b_{0} \left(\mathbf{y}_{1} \right)$ as $b \left(r \right)$ in analogy to Eq.(\ref{eq:a_2}) then:
\begin{equation}
\begin{split}
{{b}_{0}}\left( r \right)=\frac{{{b}_{2}}\left( r \right)}{r},
\label{eq:b_2}
\end{split}
\end{equation}
This change just allow us to omit the solution for the field $b_{0}\left(r \right)$, since it is completely identical to the solution for field  $a_{0} \left(r \right)$ that was written above. Afterwards we will return later to the variable $\mathbf{y}_{1}$, and perform the further computations using it. Thus, $b_{0} \left(\mathbf{y}_{1} \right)$ will have a finite value at the zero point, only if function $b_{2} \left(\mathbf{y}_{1} \right)$ will turn to zero at $\mathbf{y}_{1}=0$, which is similar to Eq.(\ref{eq:Granichni_umovi}), but the first order derivative at the zero point (value of constant $C$ in Eq.(\ref{eq:Granichni_umovi})) may be different. We denote the value of that first order derivative as $C_{1}$. From the representation in Eq.(\ref{eq:C_ne_nol}) of Eq.(\ref{eq:Dif_ur_a_2}) it follows that $C$ and $C_{1}$ are equal to $a_{0} \left(\mathbf{y}_{1} \right)$ and $b_{0} \left(\mathbf{y}_{1} \right)$ respectively at the zero value of their arguments. Therefore, in order to satisfy the requirement of Eq.(\ref{eq:umova_Jacobi}), one should use this relation:
\begin{align}
{{C}_{1}}=\frac{1}{3}C.
\label{eq:C1_C}
\end{align}

In analog to Eq.(\ref{eq:a0_plus_a1}) we get:
\begin{equation}
\begin{split}
b\left( X,\mathbf{y}_{1} \right)={{b}_{0}}\left( \mathbf{y}_{1} \right)+{{b}_{1}}\left( X, \mathbf{y}_{1} \right).
\end{split}
\label{eq:b0_plus_b1}
\end{equation}
Based on ideas used for the field $a_{1} \left(X, \mathbf{y}_{1} \right)$, one may see that the dependence of the field $b_{1} \left(X, \mathbf{y}_{1} \right)$ on the internal variable $\mathbf{y}_{1}$ should be given by the eigenfunction of the operator:
\begin{equation}
\begin{split}
& \left( \hat{H}^{\text{internal,1}} \right)^{2}   \left( b_{1}  \left( X, \mathbf{y}_{1} \right) \right) \\ 
& =-4 \Delta_{\mathbf{y}_{1}} b_{1} \left( X,\mathbf{y}_{1} \right)+{{g}^{2}}\left( -{{b}_{0}}\left( \mathbf{y}_{1} \right) \right) b_{1}  \left( X,\mathbf{y}_{1} \right), \\ 
\end{split}
\label{eq:H_vnut_kv1}
\end{equation}
where now the role of the potential of interaction is played by the function $b_{0} \left(\mathbf{y}_{1} \right)$ instead of $a_{0} \left(\mathbf{y}_{1} \right)$ as this was determined in Eq.(\ref{eq:H_vnut_kv}). In addition, the expression for both potentials will include the constant $C$, the first time as it is, at the second time through $C_{1}=C/3$.  If the internal states of the two-gluon particles that correspond to fields $a_{1} \left(X, \mathbf{y}_{1} \right)$ and $b_{1} \left(X, \mathbf{y}_{1} \right)$ will be described by eigenfunctions normalized to unity of operators in Eq.(\ref{eq:H_vnut_kv}) and Eq.(\ref{eq:H_vnut_kv1}) that in turn correspond to the lowest eigenvalues, then these functions will depend on the constant $C$. 
The normalized eigenfunctions will be denoted as $\psi_{a}(\mathbf{y}_{1})$ and $\psi_{b}(\mathbf{y}_{1})$, respectively. The condition in Eq.(\ref{eq:umova_Jacobi}) leads to the equation:
\begin{align}
{{\psi }_{a}}\left( \mathbf{y}_{1} = 0 \right)=3{{\psi }_{b}}\left( \mathbf{y}_{1} = 0 \right).
\label{eq:psia_3psib}
\end{align}
This equation determines the values of the constant $C$, hence, it also determines the boundary conditions for Eq.(\ref{eq:Dif_ur_a_2}) and for the analog equation for $b_{2} \left(\mathbf{y}_{1}\right)$.	

Let us, for instance, consider the case when $k~\textgreater~0, C~\textgreater~0$, that preserves confinement and asymptotic freedom. For this case we have the estimation in Eq.(\ref{eq:nerivist}). Thus we set approximately:
\begin{equation}
\left( -{{a}_{2}}\left( r \right) \right) \cong Cr+\frac{1}{6}\left( k+\frac{{{g}^{2}}{{C}^{2}}}{4} \right){{r}^{3}}. \\ 
\label{eq:Nablijenna_potencialu}
\end{equation}
For this approximation the properties of the eigenfunctions and eigenvalues of the operator in Eq.(\ref{eq:H_vnut_kv}) will be the same at least at the qualitative level with respect to the explicit potential. But within this approximation we obtain the precise solution for the three dimensional harmonic oscillator, since:
\begin{equation}
\left( -{{a}_{0}}\left( r \right) \right) = C+\frac{1}{6}\left( k+\frac{{{g}^{2}}{{C}^{2}}}{4} \right){{r}^{2}}. \\ 
\label{eq:Nablijenna_potencialu_a0}
\end{equation}
\begin{figure}[t]
	\centering
	\includegraphics[scale=0.32]{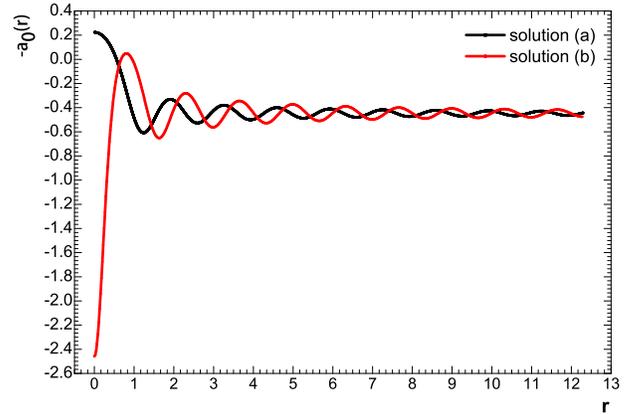}
	\caption{
		(a) Potential $-a_{0}(r)$ that corresponds to solution shown on Fig.2(a).	
		(b) Potential $-a_{0}(r)$ that corresponds to solution $-a_{2}(2)$  shown on Fig.2(c).
	}
	\label{fig:Potentials}
\end{figure}

The eigenfunction of the operator in Eq.(\ref{eq:H_vnut_kv}), normalized to unity with the help of the potential in Eq.(\ref{eq:Nablijenna_potencialu_a0}) that corresponds to the lowest eigenvalue, has the form:
\begin{equation}
\begin{split}
{{\psi }_{a}}\left( \mathbf{y}_{1} \right)={{\left( \frac{{{g}^{2}}\left( k+\frac{g^{2}C^{2}}{4} \right)}{24\pi } \right)}^{3/4}}\exp \left( -\frac{{{\left( \mathbf{y}_{1} \right)}^{2}}}{2y_{0}^{2}} \right),
\end{split}
\label{eq:Normovana_vlasna_funcia_a}
\end{equation}
where we use the notation:
\begin{align}
{{y}_{0}}=\sqrt{\frac{24}{{{g}^{2}}\left( k+ \frac{g^{2}C^{2}}{4}  \right)}}.
\label{eq:y_0}
\end{align}
For the function $\psi_{b}(\mathbf{y}_{1})$, we have similar relations but with substitution of $C/3$ instead of $C$ with respect to Eq.(\ref{eq:C1_C}). As a result of this substitution, Eq.(\ref{eq:psia_3psib}) takes form:
\begin{equation}
\begin{split}
{{\left( \frac{{{g}^{2}}\left( k + \frac{g^{2}C^{2}}{4} \right)} {24\pi } \right)}^{3/4}} = 3{{\left( \frac{{{g}^{2}}\left( k+ \frac{g^{2}C^{2}}{36} \right)}{24\pi } \right)}^{3/4}}. \\ 
\end{split}
\label{eq:Rivnanna_dla_C}
\end{equation}

Since we are considering the case when $k~\textgreater~0, C~\textgreater~0$, we are interested in a positive solution of this equation. There is only one positive solution and it has the following form:
\begin{equation}
C=\sqrt{\frac{4\left( {{3}^{{4}/{3}\;}}-1 \right)k}{{{g}^{2}}\left( 1-{{3}^{-{2}/{3}\;}} \right)}}.
\label{eq:Dodatnij_rozvazok}
\end{equation}

Taking into account the representations Eq.(\ref{eq:a0_plus_a1}) and Eq.(\ref{eq:b0_plus_b1}), by fulfilling the corresponding relations for $a_{0}\left(\mathbf{y}_{1}\right)$ and $b_{0}\left(\mathbf{y}_{1}\right)$, as well as for $a_{1}\left(X,\mathbf{y}_{1}\right)$ and $b_{1}\left(X,\mathbf{y}_{1}\right)$ we satisfied the requirement Eq.(\ref{eq:umova_Jacobi}) at zeroth order of perturbation theory for $a\left(X,\mathbf{y}_{1}\right)$ and $b\left(X,\mathbf{y}_{1}\right)$. Because we can use the representation of the interaction in which the field operators depend on their arguments in the same way as in the zeroth order approximation, we can say that the requirement, Eq.(\ref{eq:umova}), is taken into account.

By using the representation of the interaction picture\footnote{The Dirac representation}, we can define all multi-particle operators through eigenfunctions of the internal Hamiltonians or squares of internal Hamiltonians that correspond to the lowest eigenvalues. By substituting these representations into the interaction Lagrangian, one can perform integration over internal variables and obtain a model. In this model, the operators that correspond to the creation and annihilation of hadrons are coupled to operators that correspond to the creation and annihilation of two-gluon states.

\begin{figure*}[t]
	\centering
	\includegraphics[scale=0.61]{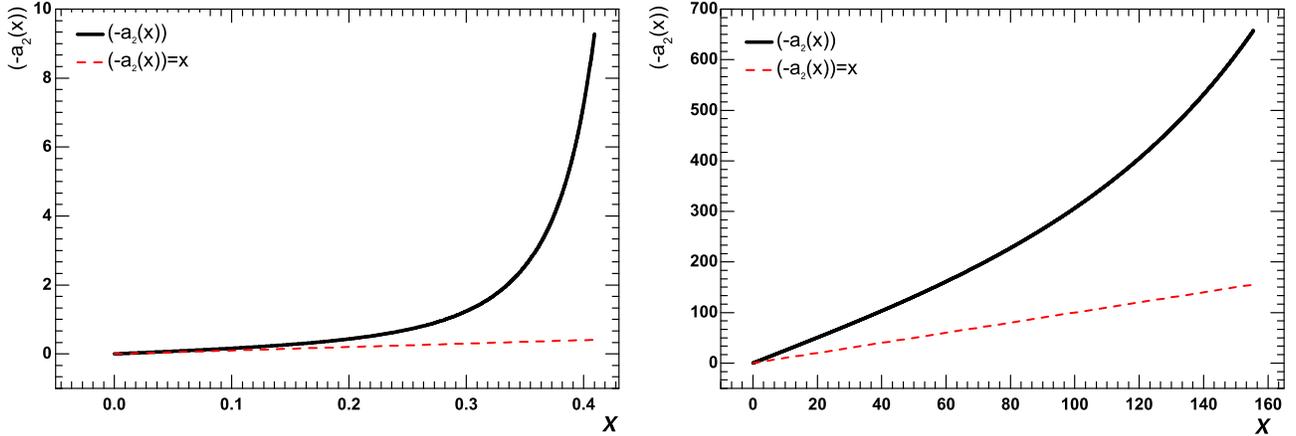}	
	\caption{
		(left) Results of the numerical solution of Eq.(\ref{eq:novij_vid}) for $C^{\prime}=1.5, g=10, k=-5$.
		(right) Results of the numerical solution of Eq.(\ref{eq:novij_vid}) for $C^{\prime}=2.5, g=0.1, k=-5$.
	}
	\label{fig:x_vs_a2x_Graph9_comb}
\end{figure*}

\section{Discussion and Conclusions}
\label{Conclusion}

The obtained results for the multi-particle fields allow one to compute quantities that are observed in experiments, for instance, in the inelastic and elastic processes of proton scattering. Indeed, we have a proton three-particle bi-spinor field that may interact with the gluon field, and this interaction is described by the Lagrangian in Eq.(\ref{eq:LintProton}). In addition, as seen from dynamical equation, see Eq.(\ref{eq:fi3_gluon}), for the two-gluon field this gluon field is self interacting. For this dynamical equation one can write a Lagrangian that turns it into an Euler-Lagrange equation. Given that Euler-Lagrange equation should include derivatives from the Lagrangian with respect to its fields, then this Lagrangian will include a Lagrangian of self interaction of cubic power, in order to provide terms of second power, see the last term in Eq.(\ref{eq:fi3_gluon}). If one will transform these multi-particle fields with respect to the representation of the interaction, then one has to substitute solutions of the equations for the free field into the Lagrangian of the interaction and self interaction. Since the dependence of the field functions of the free fields from internal variables was found by us as a solution of the eigenvalue problem for the internal Hamiltonians, there is a possibility to calculate integrals over internal variables in the interaction Lagrangian. Thus, we come to the problem, of formally matching the problem of ordinary single-particle field theory. For such a model one may use the usual methods of diagram techniques, and in this way compute differential and total cross sections that can be compared with available experimental data \cite{Antchev:2011, Aad_NatureComm:2011}. The framework shown in this paper allows one to compute the experimental properties of the inelastic scattering by adding to this model the interaction of the two-gluon field with a pseudo-scalar meson field, see the Lagrangian in Eq.(\ref{eq:L0_plus_Lint}). For both elastic and inelastic scattering we obtain the ``correct" law of the energy-momentum conservation in the sense as was discussed in the introduction. The model obtained in this way will look similar to the famous $\phi^{3}$ model. Thus the results obtained in the framework of this model \cite{cej} allow us to hope for a successful description of the experimental data.

Note, that for the reasons mentioned in the introduction, one can make the assumption that in principle the multi-particle field operators cannot be expressed through the single-particle field operators. It seems that the dynamics of multi-particle fields in principle may not be obtained from single-particle field theories. Indeed, if we consider the single-particle fields, then we come to the momentum representation starting from any kind of representations. But as a result of the requirements of a relativistic theory, the presence of the ``self" momenta in the single-particle state requires the presence of the ``self" energy. While the two-particle system only has the energy of the whole system, and not the energies of individual particles. It look like the multi-particle fields are ``independent" objects that should be considered regardless of single-particle fields.

Using this fact one can explain the difference in the potentials that provide confinement and asymptotic freedom that we found in this work with respect to those that are used in lattice calculations \cite{Simonov:1996}. Really, the lattice calculations are aimed to calculate the continiuos integral with respect to single-particle field configurations. But, as discussed in the introduction, the problem occurred before we choose the method, for instance, lattice calculations, for computing the time evolution matrix element or the operator of scattering. The problem is manifested in the choice of states for which we would like to compute the matrix element. In lattice calculations these states are constructed as result of acting with the single-particle creation operator on the vacuum. Even if one will consider the matrix element between vacuum states, then these states are defined in the way that the single-particle annihilation operator set them to zero. And integration is done over single-particle fields. Thus, if this assumption is true, then no matter how precise the calculations would be that use the single-particle fields, they are not able to take into account the multi-particle effects.





\begin{acknowledgments}
We are grateful to Jarah Evlsin for all the valuable comments and suggestions, which greatly help improve the quality of this paper. 
The work of MAD was supported by the Chinese Academy of Sciences President's International Fellowship Initiative under Grant No. 2016PM043.
\end{acknowledgments}


\begin{thebibliography}{99}
\bibliography{references}











\bibitem{SharfUJP:2011} I.V. Sharf, A.V. Tykhonov, G.O. Sokhrannyi, and others,
Description of hadron inelastic scattering by the Laplace method and new mechanisms of cross-section growth,
\textit{Ukrainian Journal of Physics} ~Vol.\textbf{56},~p.1151-1164,~2011

\bibitem{cej} I.V. Sharf, A.V. Tykhonov, G.O. Sokhrannyi, and others,
On the Role of Longitudinal Momenta in High Energy Hadron-Hadron Scattering,
\textit{Central Eur.J.Phys.} ~Vol.\textbf{10},~p.858-887,~2012, \textcolor{blue}{[arXiv:hep-ph/1110.4945]}

\bibitem{SharfP1_JModPhys:2011} I.V. Sharf, G.O. Sokhrannyi, A.V. Tykhonov and others,
Mechanisms of proton-proton inelastic cross-section  growth in multi-peripheral model within the framework of perturbation theory. Part 1,
\textit{Journal of Modern Physics} ~Vol.\textbf{2},~N\textbf{12},~p.1480-1506,~2011, \textcolor{blue}{[arXiv:hep-ph/0605110]}

\bibitem{SharfP2_JModPhys:2012} I.V. Sharf, G.O. Sokhrannyi, A.V. Tykhonov and others,
Mechanisms of proton-proton inelastic cross-section  growth in multi-peripheral model within the framework of perturbation theory. Part 2,
\textit{Journal of Modern Physics} ~Vol.\textbf{3},~N\textbf{1},~p.16-27,~2012, \textcolor{blue}{[arXiv:nucl-th/0711.3690]}

\bibitem{SharfP3_JModPhys:2012} I.V. Sharf, G.O. Sokhrannyi, A.V. Tykhonov and others,
Mechanisms of proton-proton inelastic cross-section  growth in multi-peripheral model within the framework of perturbation theory. Part 3,
\textit{Journal of Modern Physics} ~Vol.\textbf{3},~N\textbf{2},~p.129-144,~2012, \textcolor{blue}{[arXiv:hep-ph/0912.2598]}

\bibitem{Sharph:2015eka} I.V. Sharf and others,
The new method of interference contributions accounting for inelastic scattering diagrams,~2015, \textcolor{blue}{[arXiv:hep-ph/1509.04329]}

\bibitem{Sharf_Conf:2012} I.V. Sharf, K.K. Merkotan, N.A.Podolyan and others,
Gluon Loops in the Inelastic Processes in QCD,~2012, \textcolor{blue}{[arXiv:hep-ph/1210.3490]}

\bibitem{Sharf_Conf:2013} I.V. Sharf, A.V. Tykhonov, M.A. Delyiergyiev and others,
On equivalence of gluon-loop exchange in the inelastic processes in perturbative QCD to pion exchange in $\phi^{3}$ theory,
\textit{EPJ Web of Conferences} ~Vol.\textbf{60},~p.20018,~2013

\bibitem{Altland:2006} Alexander Altland and Ben Simons,
Condensed Matter Field Theory,~p.636,~2006, Cambridge University Press

\bibitem{Brodsky:1998} Stanley J. Brodsky, Hans-Christian Pauli, Stephen S. Pinsky,
Quantum chromodynamics and other field theories on the light cone,
\textit{Physics Reports} ~Vol.\textbf{301},~Issue\textbf{4-6},~p.299-486,~1998, \textcolor{blue}{[arXiv:hep-ph/9705477]}

\bibitem{Strikman:2011} Mark Strikman,
Transverse structure of the nucleon and multiparton interactions,
\textit{Prog.Theor.Phys.Suppl.} ~Vol.\textbf{187},~p.289-296,~2011, [High Energy Strong Interactions 2010: $\textendash$ Parton Distributions and Dense QCD Matter $\textendash$ Proceedings of the YIPQS International Workshop]

\bibitem{MarkusDiehl:2012} Markus Diehl, Daniel Ostermeier and Andreas Sch\"{a}fer,
Elements of a theory for multiparton interactions in QCD,
\textit{JHEP} ~Vol.\textbf{03},~p.089,~2012, \textcolor{blue}{[arXiv:hep-ph/1111.0910]}

\bibitem{GellMann1964214} M. Gell-Mann,
A schematic model of baryons and mesons,
\textit{Physics Letters} ~Vol.\textbf{08},~N\textbf{3},~p.214-215,~1964

\bibitem{Zweig:570209} G. Zweig,
An $\mathbf{SU}(3)$ model for strong interaction symmetry and its  breaking; Version 2,
\textit{CERN-TH-412} ~p.80,~1964, [Version.1 is CERN preprint 8182/TH.401, Jan. 17, 1964]

\bibitem{PhysRevLett.13.598} O.W. Greenberg,
Spin and Unitary-Spin Independence in a Paraquark Model of Baryons and Mesons,
\textit{Phys. Rev. Lett.} ~Vol.\textbf{13},~Issue \textbf{20},~p.598-602,~1964

\bibitem{PhysRev.139.B1006} M.Y. Han  and Y. Nambu,
Three-Triplet Model with Double $\mathbf{SU}(3)$ Symmetry,
\textit{Phys. Rev.} ~Vol.\textbf{139},~Issue \textbf{4B},~p.B1006-B1010,~1965


\bibitem{Merkurev:1993} S.P. Merkur'ev and L.D. Faddeev, Quantum Scattering Theory for Several Particle Systems (Mathematical Physics and Applied Mathematics),
~Vol.\textbf{11},~p.406,~1993, Springer Netherlands [Original Language: Russian]

\bibitem{SharfState:2014} I.V. Sharf, M.A. Deliyergiyev and others,
The state of non-relativistic quantum system in a relativistic reference frame, \textcolor{blue}{[arXiv:hep-ph/1403.3114]}

\bibitem{SharfState:2013} I.V. Sharf, M.A. Deliyergiyev, A.G. Kotanzhyan and others,
Transformation of the non-relativistic quantum system under transition from one inertial reference frame to another, \textcolor{blue}{[arXiv:hep-ph/1307.2280]}

\bibitem{Bogolyubov:1980} N.N. Bogolyubov and D.V. Shirkov, Introduction to the Theory of Quantized Fields,
~p.638,~1980, John Wiley \& Sons Inc, Canada Netherlands [Original Language: Russian]

\bibitem{Schrodinger:1926} E. Schr\"{o}dinger,  Quantisierung als Eigenwertproblem (Vierte Mitteilung),
\textit{Annalen der Physik} ~Vol.\textbf{81},~p.109,~1926

\bibitem{OKlein:1926} O. Klein, {Quantentheorie und f\"{u}nfdimensionale Relativit\"{a}tstheorie},
\textit{Zeitschrift f\"{u}r Physik} ~Vol.\textbf{37}, ~N\textbf{12},~p.895-906,~1926, [in German]

\bibitem{Fock:1926} V. Fock, {Zur Schr\"{o}dingerschen Wellenmechanik},
\textit{Zeitschrift f\"{u}r Physik} ~Vol.\textbf{38}, ~N\textbf{3},~p.242-250,~1926, [in German]

\bibitem{Gordon:1926} W. Gordon, {Der Comptoneffekt nach der Schr\"{o}dingerschen Theorie},
\textit{Zeitschrift f\"{u}r Physik} ~Vol.\textbf{40}, ~N\textbf{1-2},~p.117-133,~1926, [in German]

\bibitem{Nerling:2011} Frank Nerling and the COMPASS collaboration, 
Meson spectroscopy with COMPASS,
\textit{Journal of Physics: Conference Series} ~Vol.\textbf{312},~N{3},~p.032017,~2011, \textcolor{blue}{[arXiv:hep-ex/1012.0499]}

\bibitem{Lamsa:2011} J.W. L\"{a}ms\"{a} and R. Orava, 
Central diffraction at ALICE,
\textit{Journal of Instrumentation} ~Vol.\textbf{6},~N{02},~p.P02010,~2011, \textcolor{blue}{[arXiv:hep-ex/1009.3350]}

\bibitem{Frasca:2010} Marco Frasca, 
Glueball spectrum and hadronic processes in low-energy QCD, 
\textit{Nucl.Phys. B - Proceedings Supplements} ~Vol.\textbf{207–208},~p.196-199,~2010, \textcolor{blue}{[arXiv:hep-ph/1007.4479]}

\bibitem{Antchev:2011} G. Antchev and others (The TOTEM Collaboration), 
First measurement of the total proton-proton cross-section at the LHC energy of $\sqrt{s}$ = 7 TeV, 
\textit{EPL (Europhysics Letters)} ~Vol.\textbf{96},~N{2},~p.21002,~2011, \textcolor{blue}{[arXiv:hep-ex/1110.1395]}

\bibitem{Aad_NatureComm:2011} G. Aad and others (The ATLAS Collaboration), 
Measurement of the inelastic proton-proton cross-section at $\sqrt{s}$ = 7 TeV with the ATLAS detector, 
\textit{Nature Communications} ~Vol.\textbf{2},~N{463},~p.14,~2011, \textcolor{blue}{[arXiv:hep-ex/1104.0326]}

\bibitem{Simonov:1996} Yu.A. Simonov, 
The confinement, 
\textit{Physics-Uspekhi} ~Vol.\textbf{39},~N{4},~p.313-336,~1996, \textcolor{blue}{[arXiv:hep-ph/9709344]}
	





\end{thebibliography}




\end{article}








\end{document}